%% file: template.tex
\documentclass[journal]{vgtc}                     % final (journal style)
\onlineid{7364}

\vgtccategory{Research}

\title{DiffUNet$^2$: Bidirectional Conditional Diffusion for Probabilistic Scientific Spatiotemporal Modeling
}

\author{%
  \authororcid{Mengdi Chu}{0000-0003-0533-7801},
  \authororcid{Jiaxin Yang}{0009-0009-5309-4518},
  \authororcid{Angus G. Forbes}{0000-0002-8700-7795},
  \authororcid{Nathan Debardeleben}{0000-0002-5593-9205},\\
  \authororcid{Earl Lawrence}{0000-0002-6473-1887},
  \authororcid{Ayan Biswas}{0000-0002-5535-4549}, and
  \authororcid{Han-Wei Shen}{0000-0002-1211-2320}
}
\authorfooter{
  \item
  	Mengdi Chu, Jiaxin Yang, and Han-Wei Shen are with the Ohio State University.
  	E-mail: \{chu.752, yang.5039, shen.94\}@osu.edu
  \item
  	Angus G. Forbes is with NVIDIA.
  	E-mail: aforbes@nvidia.com.

  \item Nathan Debardeleben, Earl Lawrence, Ayan Biswas are with Los Alamos National Laboratory.
  	E-mail: \{ndebard, earl, ayan\}@lanl.gov.
}

\abstract{%
Studying the spatiotemporal evolution of scientific phenomena often relies on costly simulations and experiments.
Machine learning-based surrogate models reduce this cost, but most are limited to deterministic forward prediction. 
Scientific temporal analysis often requires both forward prediction and backward inference, while temporal evolution is not always uniquely determined, especially for the inverse problem.
We introduce DiffUNet$^2$, a bidirectional conditional diffusion model for probabilistic scientific temporal prediction. It supports both forward and backward prediction within a shared model.
We evaluate DiffUNet$^2$ on four scientific temporal datasets spanning
fluid dynamics, chemical reaction dynamics, and material deformation, against deterministic and probabilistic temporal baselines. Results show that DiffUNet$^2$ achieves strong predictive performance in both temporal directions and high probabilistic ensemble quality compared with existing baselines. To support practical scientific exploration beyond prediction, we further extend DiffUNet$^2$'s generation ability with target-guided state editing, allowing states of interest to be specified and explored in both temporal directions.
 The code and supplemental materials will be released upon publication.
}


\teaser{
  \centering
  \includegraphics[width=\linewidth]{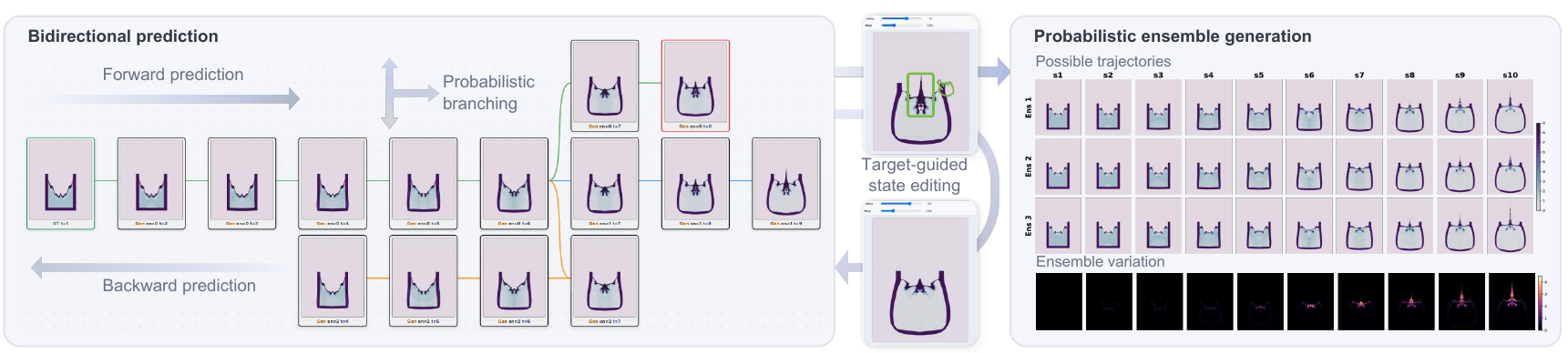}
  \caption{DiffUNet² is a bidirectional probabilistic surrogate model for scientific temporal data. It predicts plausible past and future dynamics from observed or user-specified states. By modeling ensembles of possible trajectories, it captures predictive uncertainty and enables rapid, target-guided exploration of possible causes, outcomes, and dynamics.
  }
      \label{fig:teaser}
}

\graphicspath{{figs/}{figures/}{pictures/}{images/}{./}} 
\usepackage{comment}
\usepackage{amssymb}
\usepackage{tabu}                      % only used for the table example
\usepackage{booktabs}                  % only used for the table example
\usepackage{lipsum}                    % used to generate placeholder text
\usepackage{mwe}                       % used to generate placeholder figures
\usepackage{ccicons}                   % package to be able to use icons from creative commons

\usepackage{mathptmx}                  % use matching math font
\usepackage{array}
\usepackage{makecell}
\usepackage{pifont}
\usepackage{booktabs}
\usepackage{multirow}
\usepackage{booktabs}
\usepackage{rotating}
\usepackage[table]{xcolor}
\usepackage{graphicx}
\usepackage{booktabs}
\usepackage{graphicx}
\usepackage{amsmath}
\usepackage{xcolor}

\newcolumntype{C}[1]{>{\centering\arraybackslash}m{#1}}
\newcolumntype{L}[1]{>{\raggedright\arraybackslash}m{#1}}

\begin{document}

\firstsection{Introduction}
\label{sect:intro}
\maketitle
\input{content/1_introduction}

\input{fig_model}
%\input{fig_UI}
%\input{table1_methodcompare}

\input{content/2_related_works}
\input{content/3_background}

\section{Methods}
In this section, we present the formulation, architecture, and training objectives of DiffUNet$^2$. We then describe extensions designed to enhance the model’s practical utility for scientific exploration.
%As described in \autoref{sec:ChallengeGoals}, these scientific temporal data analysis challenges span three key stages: data ambiguity diagnosis, temporal evolution exploration and hypothesis testing, and comparison of plausible outcomes. To address them, we propose the framework shown in ~\autoref{fig:teaser} to support real-world scientific analysis workflows.
%Our framework has two integrated components: a bidirectional diffusion model (\autoref{fig:model}) and an interactive visual system (\autoref{fig:UI})
%Guided by the design goals, the model is designed to provide a generative backbone for scientific temporal data: accurate prediction when the evolution is deterministic (\textbf{\textit{M1}}), probabilistic ensemble generation when multiple outcomes are plausible (\textbf{\textit{M2}}), flexible generation across time for both forward and backward reasoning (\textbf{\textit{M3}}), and user-guided exploration of the data distribution by refining edited states into nearby plausible configurations (\textbf{\textit{M4}}).
%The interactive system turns these generative capabilities into visual analysis operations that support uncertainty diagnosis, test hypotheses through interactive editing (\textbf{\textit{S1-2}}), and explore multiple plausible outcomes in an intuitive workflow (\textbf{\textit{S3}}).

\subsection{Problem Formulation}
%Scientific spatiotemporal data represent the temporal evolution of a physical system through a sequence of spatial states. 
Scientific spatiotemporal data describe the evolution of a physical system through a sequence of spatial states over time. Let $\mathcal{X} = \{x_{s_0}, x_{s_1}, \ldots, x_{s_{T-1}}\}$ where $x_{s_i}$ represents the system state at time $s_i$. 
Depending on the scientific process, the temporal evolution may also be associated with physical or simulation parameters $m$. Here, we consider temporal prediction as a conditional query between system states at different time points.
Given a state $x_{s_i}$ at time $s_i$, the goal is to infer the system state at a queried time $s_j$. 
Both forward and backward temporal queries are supported within the same modeling framework.
However, the relation between states across time is not always deterministic or invertible.
When the available state does not fully determine the dynamics, multiple states may be plausible. Backward inference can be more underdetermined, particularly in ill-posed settings where multiple antecedent states may lead to a similar later observation.
Additionally, insufficient information, measurement limitations, and stochastic dynamics in practical modeling can further increase this ambiguity. 
Therefore, we model a temporal query as a conditional distribution rather than a deterministic mapping.
%Given a conditioning state $x_{s_i}$ with an optional parameter $m$, and a queried time $s_j$, we model the distribution
\begin{equation}
p_{\theta}\left(
x_{s_j}
\mid
x_{s_i}, s_i, s_j, m
\right).
\label{eq:state_query}
\end{equation}
Here, $x_{s_i}$ provides the known spatial context, $(s_i,s_j)$ specifies the temporal query, and $m$ adds physical constraints when such information is available. 
When $s_j > s_i$, the query corresponds to forward temporal prediction. 
When $s_j < s_i$, it corresponds to backward inference of history states. 
%Both forward and backward temporal queries are supported within the same modeling framework.
DiffUNet$^2$ models Eq.~\ref{eq:state_query} using a conditional residual diffusion model shared across forward and backward temporal queries.
We next describe the model formulation, bidirectional temporal supervision, and the ambiguity-preserving cycle objective.
Finally, we introduce practical extensions that improve the model's applicability to scientific exploration, including adaptations to different data representations, target-guided state editing, and an interactive interface.

\begin{figure}[t]
    \centering
   % \vspace{-8pt}
    \includegraphics[width=\linewidth]{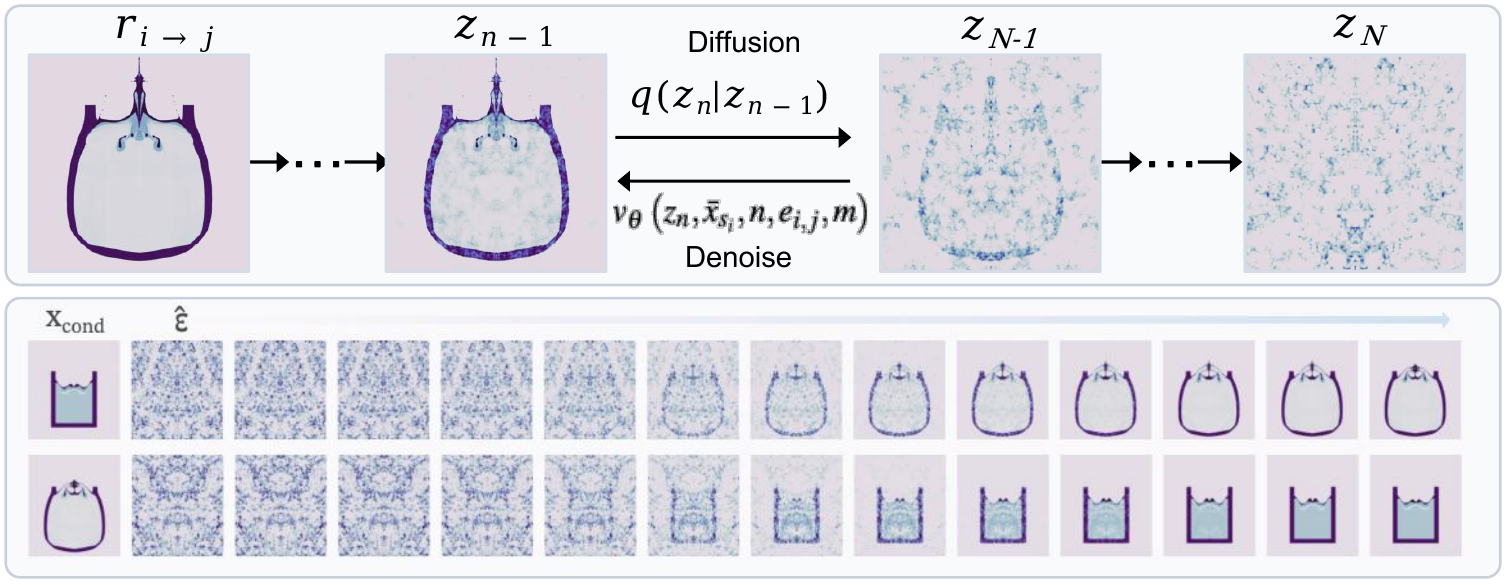}
    \caption{Conditional residual diffusion in DiffUNet$^2$. Temporal residual $r_{i\rightarrow j}$ is diffused to Gaussian noise and recovered through conditional denoising. The predicted residual is rescaled and added to the source state to obtain the target state.
    }
   % \vspace{-10pt}
    \label{fig:diffusionprocess}
\end{figure}

\subsection{Conditional Residual Diffusion}

For each temporal query $(i,j)$, DiffUNet$^2$ models the transition from the conditioning state $x_{s_i}$ to the target state $x_{s_j}$ in residual space.
The diffusion target is defined by the state change relative to the conditioning state $x_{s_i}$.
All field channels are normalized using statistics computed from the training set. For a directed pair $(i,j)$, the standardized residual is
$r_{i\rightarrow j}
=
(\bar{x}_{s_j}-\bar{x}_{s_i})/\gamma_{i,j}$,
where $\bar{x}$ denotes the normalized state and $\gamma_{i,j}$ denotes the per-channel root-mean-square residual magnitude estimated from the training
set.
At diffusion step $n$, Gaussian noise
$\epsilon \sim \mathcal{N}(0,I)$ is applied to the residual:
\begin{equation}
z_n =
\sqrt{\bar{\alpha}_n} r_{i\rightarrow j}
+
\sqrt{1-\bar{\alpha}_n}\epsilon,
\qquad
v_n^{*} =
\sqrt{\bar{\alpha}_n}\epsilon
-
\sqrt{1-\bar{\alpha}_n}r_{i\rightarrow j}.
\label{eq:forward_diffusion}
\end{equation}
We use velocity prediction and optimize
\begin{equation}
\mathcal{L}_{\mathrm{diff}}
=
\mathbb{E}_{(i,j),n,\epsilon}
\left[
\left\|
v_{\theta}
\left(
z_n,
\bar{x}_{s_i},
n,
e_{i,j},
m
\right)
-
v_n^{*}
\right\|_2^2
\right],
\label{eq:diffusion_loss}
\end{equation}
where $e_{i,j}$ encodes the directed temporal query and $m$ denotes optional parameters.
The noisy residual $z_n$ and conditioning state $\bar{x}_{s_i}$ are concatenated along the channel dimension and processed by a three-level
encoder--decoder U-Net with channel multipliers $(1,2,4)$. The diffusion-step embedding and temporal-query embedding $e_{i,j}$ are injected into the residual blocks through FiLM modulation. A single denoising network is shared across forward and backward temporal queries.
At inference time, sampling starts from
$z_N \sim \mathcal{N}(0,I)$ and proceeds through deterministic DDIM updates
to obtain $\hat{r}_{i\rightarrow j}$. The target state is recovered as
\begin{equation}
\hat{\bar{x}}_{s_j}
=
\bar{x}_{s_i}
+
\gamma_{i,j}\hat{r}_{i\rightarrow j},
\label{eq:state_recovery}
\end{equation}
followed by inverse normalization. We maintain an exponential moving average
of the denoiser for sampling. Additional sampling and reproducibility settings
are provided in Appendix~B.2.

\subsection{Bidirectional Temporal Supervision}
\label{Bidirectional Temporal Supervision}
Scientific temporal prediction is commonly formulated as either
autoregressive rollout or direct prediction to a queried time.
Autoregressive prediction learns local temporal transitions and repeatedly
applies them to generate a trajectory, while direct prediction maps an observed
state to a target time without generating intermediate states.
We include both settings with bidirectional supervision as DiffUNet$^2$-AR and DiffUNet$^2$-DR.
For both variants, each selected temporal relation is included in both
directions, allowing the same denoiser to learn increasing- and decreasing-time
transitions.
The temporal query embedding $e_{i,j}$ specifies the source time, target time,
and prediction direction.

DiffUNet$^2$-AR is trained on adjacent temporal transitions in both directions.
Each state is used to predict its neighboring state forward and backward in
time.
At inference, the model is repeatedly applied to generate a trajectory from
the observed state.

DiffUNet$^2$-DR is trained on direct transitions between sequence endpoints
and the remaining sampled time points, together with their reverse
transitions. Each queried state is therefore generated directly from the observed endpoint
without using intermediate predictions.
The query embedding additionally encodes the temporal interval so that the
same model can distinguish different source--target pairs.

Both variants use the same conditional residual diffusion model and differ
only in the temporal pairs used for supervision.
DiffUNet$^2$-AR learns local transitions for sequential rollout, whereas
DiffUNet$^2$-DR learns direct transitions between distant temporal states.

\begin{figure}[t]
    \centering
    \includegraphics[width=\columnwidth]{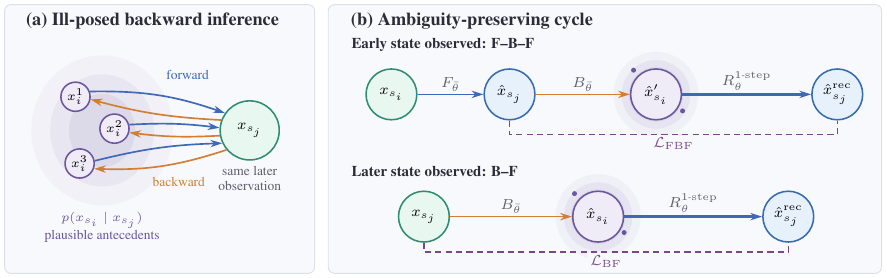}
    \caption{Ambiguity-preserving cycle consistency for backward inference.
    (a) Multiple plausible earlier states may correspond to the same later observation, making backward inference non-unique. (b) The proposed cycle objective constrains generated antecedents through reconstruction of the later state. For an observed early state, we use a forward--backward--forward (F--B--F) path; for an observed later state, we use a backward--forward (B--F) path. Consistency is imposed only in the later-state space, so generated antecedents are not forced to match a single recorded history.
    }
    \label{fig:asymmetric-cycle}
\end{figure}

\subsection{Ambiguity-Preserving Cycle Consistency}
\label{Ambiguity-Preserving Cycle Consistency}
Bidirectional supervision trains forward and backward transitions jointly but does not explicitly constrain their consistency across directions.
A standard cycle objective maps a state forward and then backward and requires the recovered state to match the original one.
However, for scientific temporal dynamics, backward inference often takes the form of a non-unique inverse problem, where multiple antecedent states may be consistent with the same later observation. Requiring cycle reconstruction to match a single recorded history can suppress plausible alternatives.
Therefore, we impose consistency in the later-state space. Given a backward-generated state, we require a forward prediction
from this state returns to the later state that it is intended to explain.
This preserves multiple plausible backward solutions while constraining them
to remain compatible with the learned temporal dynamics.

Forward--backward--forward consistency.
Starting from an observed early state $x_{s_i}$, we first generate a later state
and then infer an antecedent,
\begin{equation}
\hat{x}_{s_j}
=
G_{\bar{\theta}}(x_{s_i}, i\rightarrow j; \xi_1),
\qquad
\hat{x}^{0}_{s_i}
=
G_{\bar{\theta}}(\hat{x}_{s_j}, j\rightarrow i; \xi_2),
\end{equation}
where $G_{\bar{\theta}}$ denotes DDIM sampling with the EMA model.
The generated antecedent is then used to predict the later state again, and
the cycle loss is computed in the later-state space.
\begin{equation}
\mathcal{L}_{\mathrm{FBF}}
=
\left\|
R_{\theta}
\left(
\hat{x}^{0}_{s_i},
\operatorname{sg}(\hat{x}_{s_j}),
i\rightarrow j
\right)
-
\operatorname{sg}(\hat{x}_{s_j})
\right\|_2^2.
\label{eq:fbf_cycle}
\end{equation}

Backward--forward consistency. When the later state $x_{s_j}$ is observed, we first generate a plausible
antecedent, and require it to reproduce the observed later state.
\begin{equation}
\hat{x}_{s_i}
=
G_{\bar{\theta}}(x_{s_j}, j\!\rightarrow\! i; \xi),
\qquad
\mathcal{L}_{\mathrm{BF}}
=
\left\|
R_{\theta}(\hat{x}_{s_i},x_{s_j},i\!\rightarrow\! j)
-
x_{s_j}
\right\|_2^2.
\label{eq:bf_cycle}
\end{equation}

Here $R_{\theta}$ denotes a single-step clean-state estimate from the current denoiser.
The intermediate states are generated with the detached EMA model, while gradients are propagated only through $R_{\theta}$.
The complete training objective is
$\mathcal{L}=\mathcal{L}_{\mathrm{diff}}+\lambda_{\mathrm{cyc}}\mathcal{L}_{\mathrm{cyc}}$,
where $\mathcal{L}_{\mathrm{cyc}}$ denotes the corresponding FBF or BF consistency term.

\subsection{Extensions for Practical Scientific Exploration}
We further extend DiffUNet$^2$ to support practical scientific data exploration. These extensions include adaptations to physical-, latent-, and volumetric data representations, together with target-guided state editing and an interface for interactive bidirectional temporal exploration.

\begin{figure}[t]
    \centering
    \vspace{-8pt}
    \includegraphics[width=\linewidth]{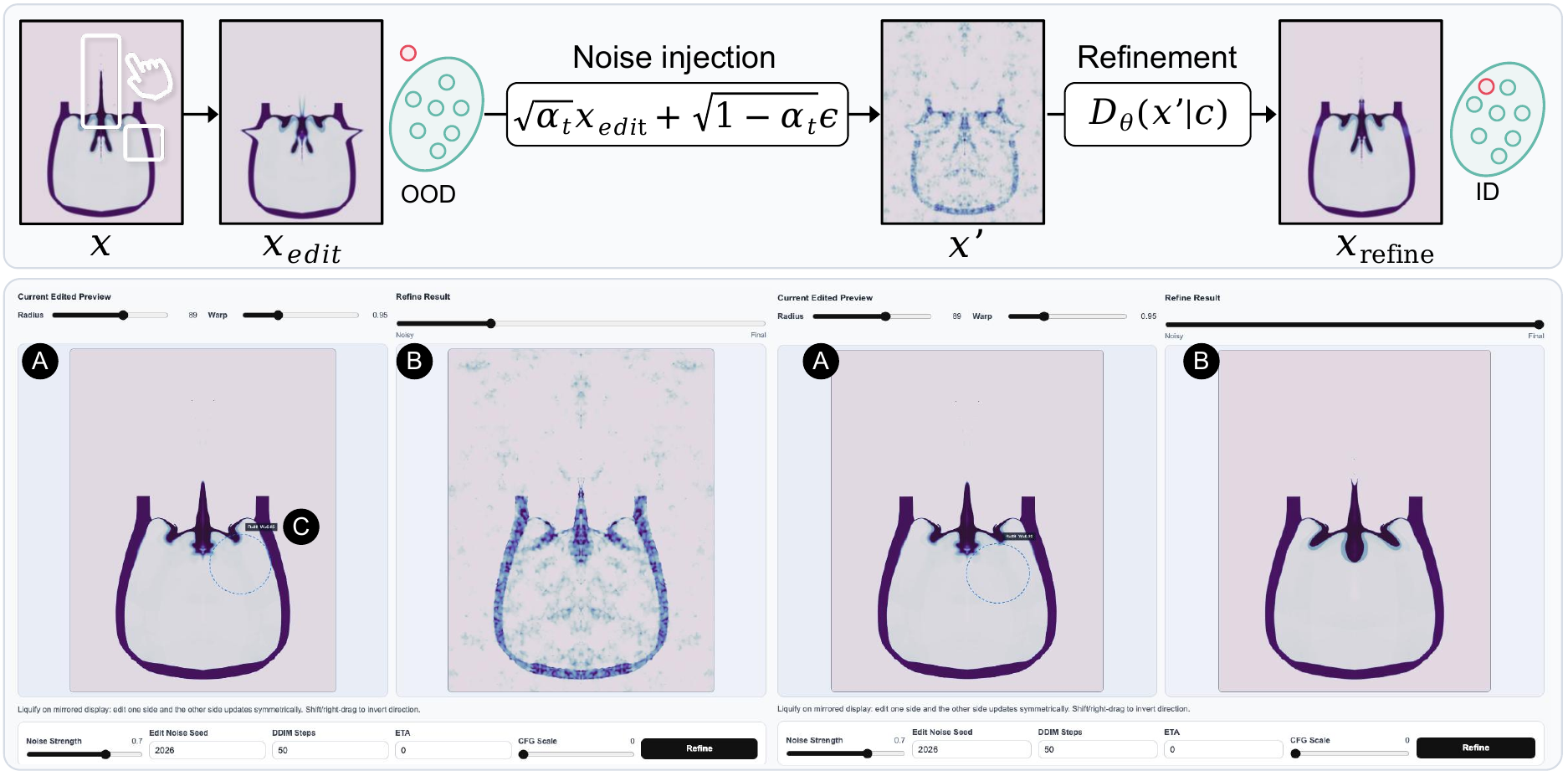}
    \caption{ Target-guided editing and diffusion refinement. Users can directly edit generated samples toward desired outcomes, while the refinement stage corrects implausible or out-of-distribution (OOD) patterns to in-distribution (ID) patterns and preserves the intended intervention.
    State editing and refinement interface. (A) Editable field view, where users modify the state directly in the spatial domain. (B) Distribution-aware refinement result generated from the edited state. (C) Brush interaction used to locally adjust structures in the field.
    }
    %\vspace{-10pt}
    \label{fig:refin}
\end{figure}

\textbf{Target-guided state generation and editing.}
\label{Target-guided state generation and editing.}
Most temporal prediction models generate sequences from fixed inputs, providing limited support for modifying system states and exploring how such interventions affect temporal evolution. 
In scientific workflows, obtaining a state with a desired structure often requires costly repeated simulation under different initial conditions or physical parameters. 
To support this form of exploration, DiffUNet$^2$ extends the diffusion formulation to target-guided state generation and editing. 
By removing temporal and parameter conditioning, the model learns the distribution of the whole data space. Users can then edit a state toward a desired configuration and refine the edited result through partial diffusion denoising.
A user can modify a state $x$ to obtain an edited target $x_{\mathrm{edit}}$ that reflects a desired configuration. Free-form edits move the state away from the learned data distribution, so we refine the edited state through partial diffusion denoising. At a selected diffusion step $n$, the edited state is perturbed as
$x_n=\sqrt{\bar{\alpha}_n}x_{\mathrm{edit}}
+\sqrt{1-\bar{\alpha}_n}\epsilon$,
where $\epsilon\sim\mathcal{N}(0,I)$.
Starting from $x_n$, the reverse diffusion process produces a refined state $x_{\mathrm{refine}}$ that preserves the intended edit while moving the sample toward the learned state distribution. The diffusion step $n$ controls the degree of refinement.
The refined state can subsequently be used as a target or conditioning state for bidirectional temporal queries, allowing the model to infer its initial condition and possible evolution (\autoref{fig:refin}).

\begin{figure}[t]
    \centering
    \vspace{-8pt}
    \includegraphics[width=\linewidth]{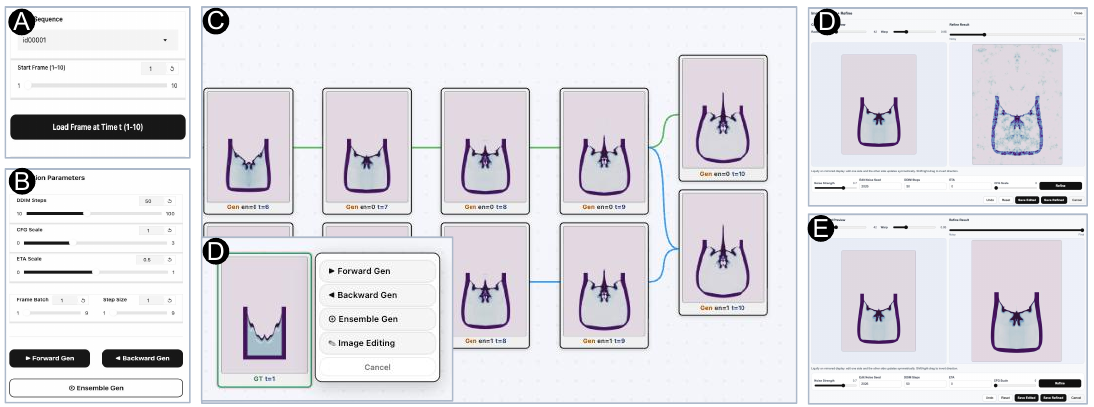}
    \caption{ Interface for interactive bidirectional exploration with DiffUNet$^2$. (A--B) Users select temporal conditions and control forward, backward, and ensemble generation.
(C) Generated states are organized as branching trajectories, and any state
can be reused as a new temporal condition.
(D) Node actions support further generation or direct state editing.
(E--F) Edited states are refined through diffusion toward the learned data
distribution and can then be used for continued bidirectional exploration.
}
    %\vspace{-10pt}
    \label{fig:interface}
\end{figure}

\textbf{Interactive temporal exploration interface}.
We integrate the model into an interface for interactive temporal exploration (\autoref{fig:interface}). Each observed, generated, or edited state is represented as a node at its corresponding time step. From any selected node, users can perform bidiractional queries, generate multiple plausible states, or edit the current state to create a new exploration branch.  Generated and
refined states can be reused as new conditions, allowing the exploration to continue across time rather than ending with a single prediction.

\textbf{Adaptation to different data representations.}
To accommodate scientific data with different spatial resolutions and representations, we adapt the same conditional residual diffusion formulation to physical and latent spaces.
For moderate-resolution 2D fields, the diffusion process operates directly in the physical data space using a 2D U-Net.
For high-resolution 2D fields, we use a KL-regularized variational autoencoder to map the data into a compact latent representation, and diffusion is performed on this latent residual. Generated latent states are decoded back to the physical space for analysis and evaluation.
The same formulation can also be extended to volumetric fields by replacing the spatial operations with their 3D counterparts, including 3D convolutions and volumetric upsampling.
The objective and temporal conditioning remain unchanged, while the network architecture is adapted and trained separately for each data representation.
%For volumetric data, we retain the same residual diffusion formulation and replace the spatial operations with 3-D Unet. The physical-space 2D, latent-space, and volumetric variants follow the same modeling formulation, with separately trained networks adapted to their spatial representations.

\section{Evaluation and Results}

We evaluate DiffUNet$^2$ across multiple scientific domains and compare it with recent deterministic and probabilistic temporal surrogate models.
The evaluation examines prediction accuracy in both temporal directions and probabilistic ensemble quality. The main comparison uses four 2D datasets, with an additional volumetric
dataset used to examine the extension to native 3D fields.

\subsection{Experimental Setup}

\textbf{Datasets.}
We select datasets from different scientific domains and with different temporal dynamics.
The main evaluation includes CERM~\cite{herde2024poseidon},
HEAT-PLI~\cite{Banesh2025HEATPLI},
Gray--Scott~\cite{pearson1993patterns}, and
NS-SVS~\cite{herde2024poseidon}.
They cover compressible and incompressible fluid dynamics,
reaction--diffusion processes, and shock-driven material deformation.
CERM, Gray--Scott, and NS-SVS are modeled directly in the physical space. For the high-resolution HEAT-PLI data, we perform temporal modeling in a compact latent space using VAE.
We additionally include 3D Kolmogorov Flow as a volumetric test case to examine the applicability of the model to native 3D fields.
\autoref{tab:eval-datasets} summarizes the predicted fields, spatial resolutions,
trajectory counts, sequence lengths, and temporal ranges.

\begin{itemize}
    \item \textbf{CERM}~\cite{herde2024poseidon} models Richtmyer--Meshkov instability governed by the compressible Euler equations. A shock interacts with a fluid interface and drives deformation and mixing over time. We predict density, two velocity components, and pressure. The forward evolution is deterministic, while shock interaction and mixing can make backward recovery less constrained.

    \item \textbf{Gray--Scott}~\cite{pearson1993patterns} is a two-species reaction--diffusion system in which nonlinear reactions and diffusion produce evolving spatial patterns. We generate the trajectories using APEBench~\cite{koehler2024apebench}, with fixed reaction parameters and varying initial states. Although the forward dynamics are deterministic, diffusion can reduce information about earlier spatial configurations and make backward prediction less constrained.

    \item \textbf{HEAT-PLI}~\cite{Banesh2025HEATPLI} is the Perturbed Layer Interface subset of the High Explosives and Affected Targets dataset. It models shock-driven deformation of cylindrically symmetric multi-material structures under different initial interface geometries. We use the density field at a spatial resolution of $1120\times400$. For efficient modeling, it is encoded into a four-channel $140\times50$ latent representation, and generated states are decoded back to the physical space for evaluation. Only density is used for temporal prediction, while the complete simulation
contains additional physical variables. This incomplete state information makes the temporal relation less constrained, especially in the backward direction.

    \item \textbf{NS-SVS}~\cite{herde2024poseidon} models incompressible Navier--Stokes flow initialized with sinusoidal vortex sheets. We predict two velocity components and a passive tracer. The dynamics contain vortex evolution, advection, and mixing. Viscous dissipation can remove fine-scale information over time, making earlier states harder to recover from later observations.

    \item \textbf{Kolmogorov Flow}~\cite{koehler2024apebench} is a periodically forced incompressible flow with nonlinear advection and viscous dissipation. We generate the trajectories using APEBench and predict the three velocity components on a $64^3$ grid. This dataset is used separately to evaluate native 3D forward and backward prediction.
\end{itemize}

\input{table/table_data}

\begin{figure}[t]
    \centering
    \includegraphics[width=\columnwidth]{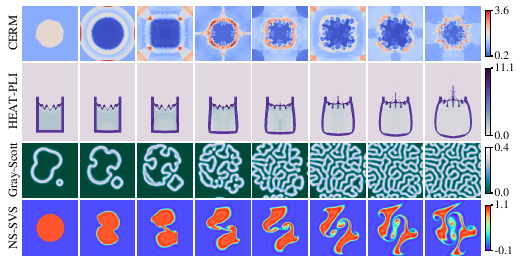}
    \caption{Example trajectories from the four 2D datasets. The forward prediction task conditions on the first frame to predict subsequent states, while the backward prediction task conditions on
the last frame to infer preceding states.}
    \label{fig:dataset_examples}
\end{figure}

\begin{figure*}[t]
    \centering
    \includegraphics[width=\textwidth]{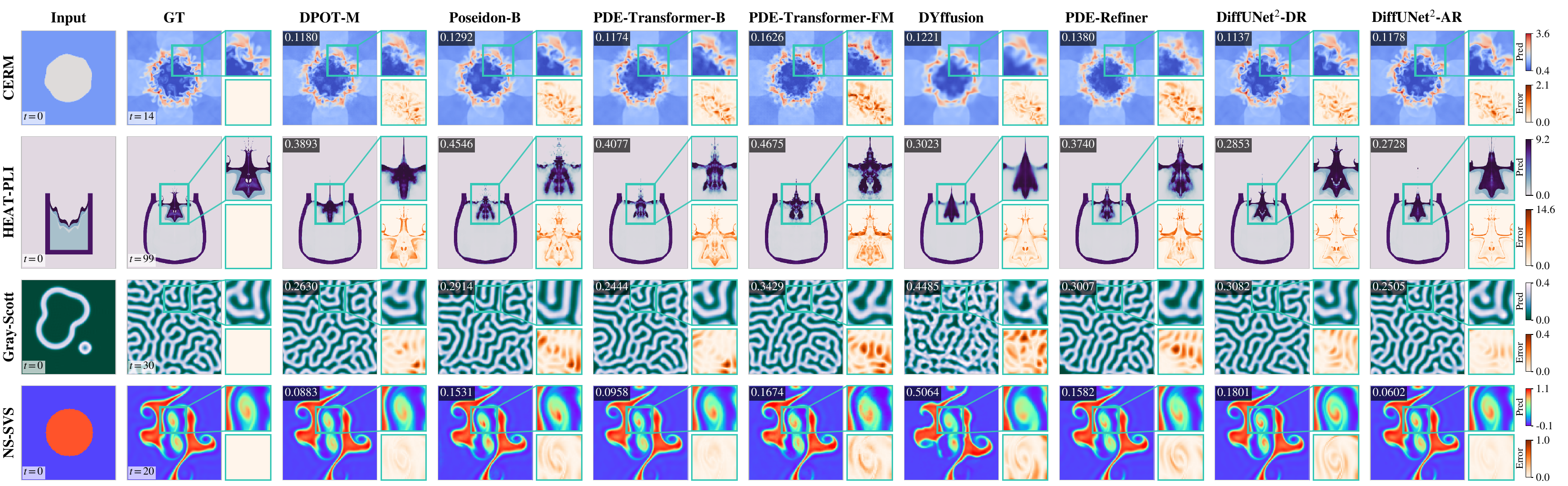}
    \caption{Forward prediction on four datasets, conditioned on
    the first frame and evaluated at the last frame. 
For probabilistic models, we display the middle-ranked sample among 32 samples ordered by nRMSE.
The outlined region is enlarged to highlight local spatial structures, with the
corresponding absolute-error map shown below.
White numbers report nRMSE for the displayed prediction.}
    \label{fig:forward_qualitative}
\end{figure*}

\begin{figure*}[h]
    \centering
    \includegraphics[width=\textwidth]{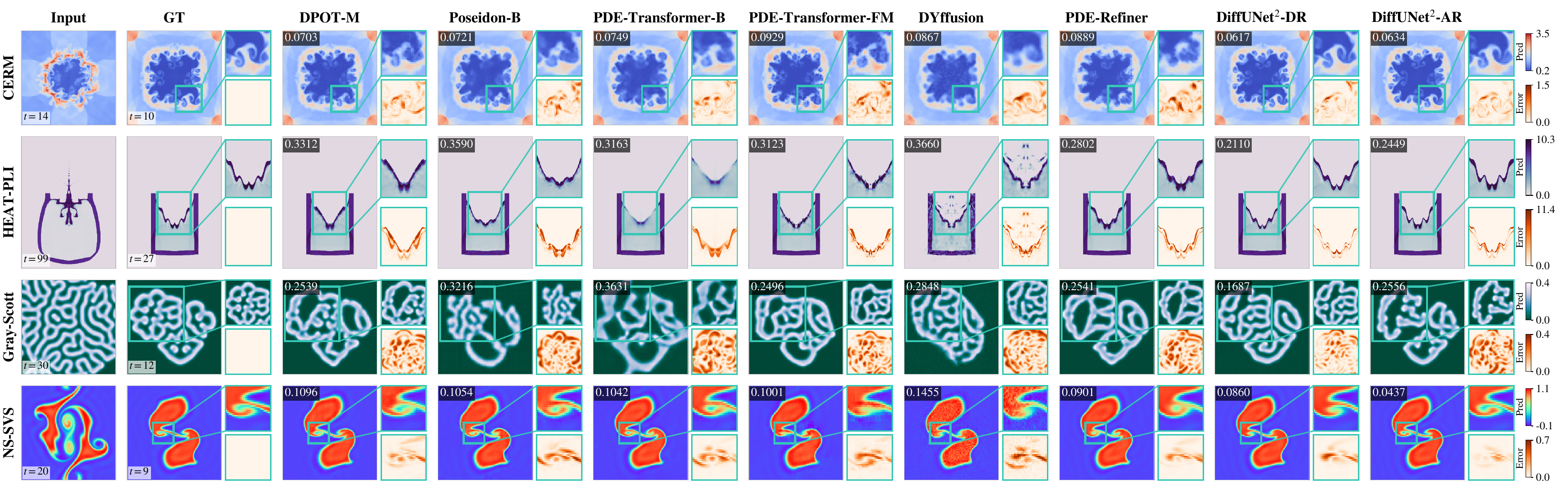}
    \caption{Backward prediction results on four 2D datasets.
Each model is conditioned on the last frame and predicts an earlier queried
state. For probabilistic models, we display the middle-ranked sample among 32 generated samples ordered by nRMSE}
    \label{fig:backward_qualitative}
\end{figure*}

\input{table/table1}

\input{table/table_modelcompare}
\input{table/table2}

\textbf{Baselines}. We compare DiffUNet$^2$ with six recent competitive models for scientific
temporal prediction, including three deterministic and three probabilistic
baselines (\autoref{tab:eval-baselines}).

For deterministic prediction, we use DPOT-M~\cite{hao2024dpot}, Poseidon-B~\cite{herde2024poseidon}, and
PDE-Transformer-B~\cite{holzschuh2025pdetransformer}. DPOT-M uses a pretrained operator backbone for autoregressive PDE prediction, Poseidon-B uses a multiscale operator transformer pretrained across
time-dependent PDEs, and PDE-Transformer-B applies a transformer architecture to temporal prediction of physical fields. These models represent recent developments in learned scientific surrogate modeling and provide strong deterministic references for prediction accuracy.

For probabilistic prediction, we compare with PDE-Transformer FM~\cite{holzschuh2025pdetransformer,lipman2023flowmatching},
PDE-Refiner~\cite{lippe2023pderefiner}, and DYffusion~\cite{cachay2023dyffusion}. PDE-Transformer-FM is the flow-matching variant of PDE-Transformer, PDE-Refiner uses iterative denoising to refine PDE predictions, and DYffusion combines temporal interpolation with diffusion-based forecasting. Together, these baselines cover representative deterministic, diffusion-based,
and flow-based approaches to scientific temporal prediction.

\textbf{Prediction task.}
All 6 baseline models are originally designed for unidirectional temporal
prediction.
We therefore train separate models for forward and backward prediction.
For forward training, each trajectory follows its original temporal order,
while the sequence is reversed for backward training.
In contrast, DiffUNet$^2$-AR and DiffUNet$^2$-DR each use a single model
jointly trained on both temporal directions.
For forward prediction, the first frame is used as the observed condition and the remaining states are predicted at later time points. For backward prediction, the last frame is used as the condition and the preceding states are predicted in reverse temporal order.
Autoregressive models use their own predictions as inputs for subsequent steps, while direct models predict each queried state directly from the observed endpoint.
Within each dataset, all models use the same trajectory-level training, validation, and test splits, spatial resolution, selected temporal states, and prediction horizons.
All predictions are evaluated in the physical space using the same metric
implementations.

\textbf{Implementation settings.}
All experiments are conducted on NVIDIA A100 GPUs with 80\,GB of memory.
For the 2D experiments, we use an effective batch size of 64 with gradient accumulation when needed and train each model for up to 500 epochs. Validation is performed every five epochs, and early stopping is applied after epoch 200 based on validation performance. All checkpoints are selected using validation data only. DiffUNet$^2$ is optimized using AdamW with a learning rate of $2\times10^{-4}$ and a weight decay of $10^{-5}$. We use 1,000 diffusion steps during training and 50-step DDIM sampling with
EMA weights at inference. Additional model-specific training and sampling settings are provided in the
supplementary material.

\autoref{tab:eval-baselines} reports model size and computational cost.
Training time is averaged over the four 2D datasets.
For direction-specific baselines, the reported training time includes both
forward and backward models, while DiffUNet$^2$ uses one jointly trained model.

\textbf{Evaluation metrics.}
We evaluate prediction accuracy using normalized root mean square error
(nRMSE), structural similarity (SSIM), and peak signal-to-noise ratio (PSNR).
For probabilistic models, we generate 32 samples for each query and use the
ensemble mean for point-prediction metrics. We further evaluate the generated ensembles using the continuous ranked probability score (CRPS) and the spread--skill ratio (SSR). CRPS measures the agreement between the predicted distribution and the observed target. SSR compares ensemble spread with prediction error. All metrics are computed in the physical space.

\begin{figure}[t]
    \centering
    \includegraphics[width=0.49\textwidth]{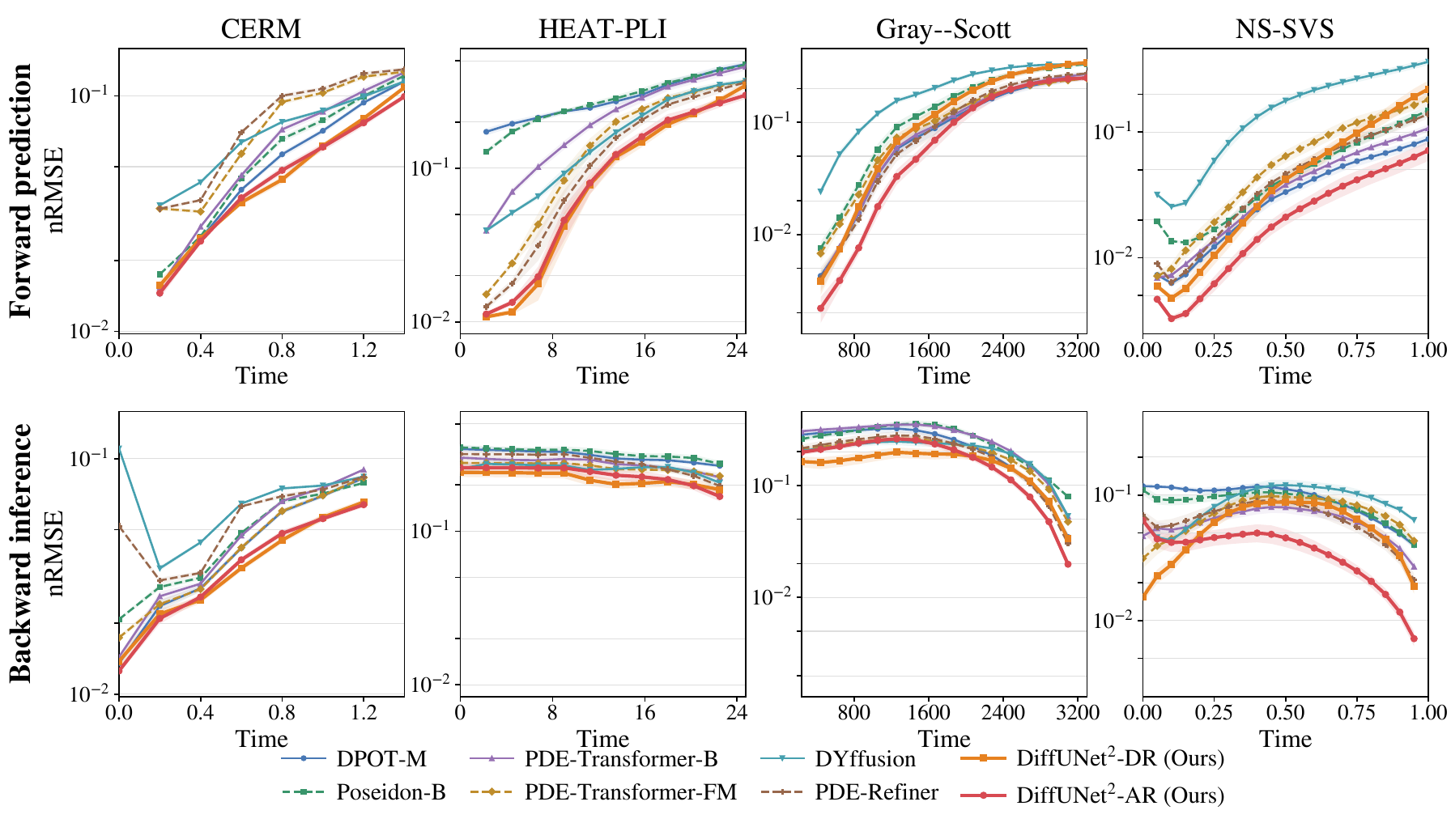}
    \caption{Temporal prediction errors on four 2D datasets.
    The top and bottom rows show forward prediction conditioned on
    the first frame and backward prediction conditioned on the last
    frame. Both rows use physical simulation time, with matching axis ranges for each dataset.
    Curves show nRMSE on a logarithmic scale; shaded bands indicate
    pointwise 95\% bootstrap confidence intervals over test trajectories.
    Probabilistic models use the mean of 32 generated
    samples. }
    \label{fig:temporal_nrmse}
\end{figure}

\subsection{Bidirectional Prediction Accuracy}

We first compare forward and backward prediction accuracy with baseline models.
\autoref{tab:main_prediction_accuracy} shows that DiffUNet$^2$ remains competitive in both temporal directions while using one shared model.
DiffUNet$^2$-AR achieves the lowest forward nRMSE on all four datasets. DiffUNet$^2$-DR performs best for backward prediction on CERM, HEAT-PLI, and Gray--Scott. On NS-SVS, AR performs best in both directions.

For forward prediction, DiffUNet$^2$-AR achieves the lowest nRMSE on all four
2D datasets.
On CERM, its nRMSE is 0.0572, compared with 0.0668 for DPOT-M. On HEAT-PLI, DiffUNet$^2$-AR reduces nRMSE to 0.1639, compared with 0.2037 for PDE-Refiner.
On Gray--Scott, the difference is smaller, with 0.1287 compared with 0.1298
for DPOT-M.
On NS-SVS, DiffUNet$^2$-AR achieves 0.0337 nRMSE, compared with 0.0457 for
DPOT-M. SSIM and PSNR show the same overall trend.
For backward prediction, DiffUNet$^2$ also shows clear advantages across the
four datasets.
DiffUNet$^2$-DR gives the lowest nRMSE on CERM, HEAT-PLI, and
Gray--Scott, while DiffUNet$^2$-AR performs best on NS-SVS.
SSIM and PSNR show the same pattern, with DR performing best on the
first three datasets and AR on NS-SVS.

The qualitative comparisons in ~\autoref{fig:forward_qualitative} and
\autoref{fig:backward_qualitative} show the same overall behavior.
DiffUNet$^2$ preserves the details interfaces, material boundaries, and evolving spatial structures while producing smaller local errors in many challenging regions.
\autoref{fig:temporal_nrmse} further shows that both DiffUNet$^2$ variants maintain competitive performance over time for both forward and backward prediction across all four datasets.
%Overall, DiffUNet$^2$ achieves strong prediction accuracy against both deterministic and probabilistic baselines in forward and backward directions, while using a single shared model for both temporal queries.
As shown in \autoref{tab:eval-baselines}, DiffUNet$^2$ 
contains about 57M parameters, fewer than most operator- and
transformer-based baselines. Although DiffUNet$^2$ performs probabilistic sampling, its training and evaluation costs remain comparable to other probabilistic models.
DiffUNet$^2$ achieves strong bidirectional prediction accuracy with a moderate model size and computational cost, while using a single shared model for both temporal directions.

%DiffUNet$^2$-AR is consistently strong for forward rollout, where local
%transitions are repeatedly composed over time.
%DiffUNet$^2$-DR performs better on several backward tasks, where the queried earlier state is generated directly from the later observation without intermediate rollout. The results show that a shared bidirectional model can achieve competitive or better prediction accuracy than direction-specific baselines, while the preferred temporal supervision depends on the dataset and prediction direction.

\begin{figure*}[t]
    \centering
    \includegraphics[width=0.495\textwidth]{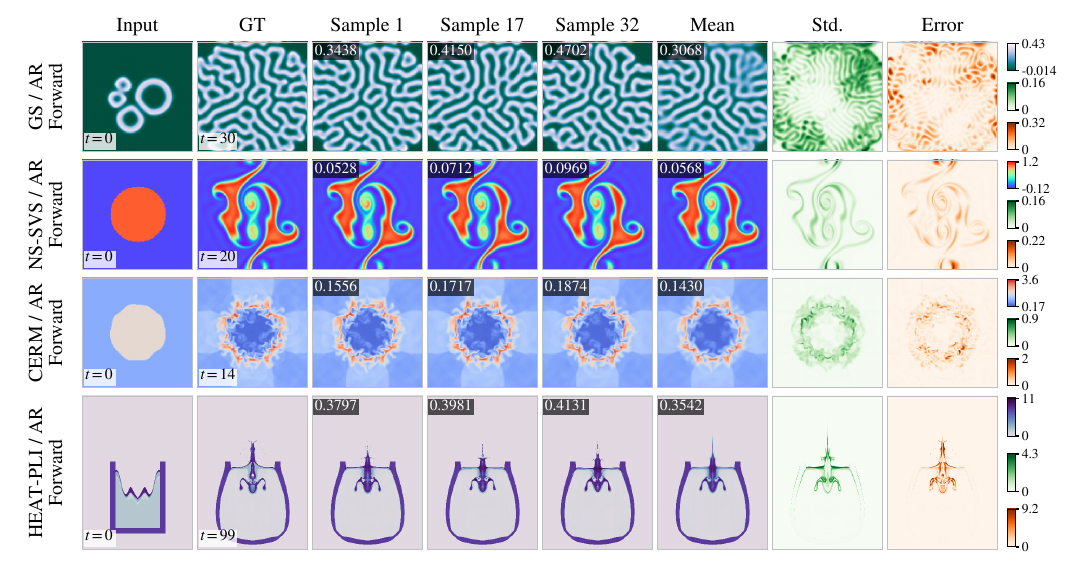}
    \includegraphics[width=0.495\textwidth]{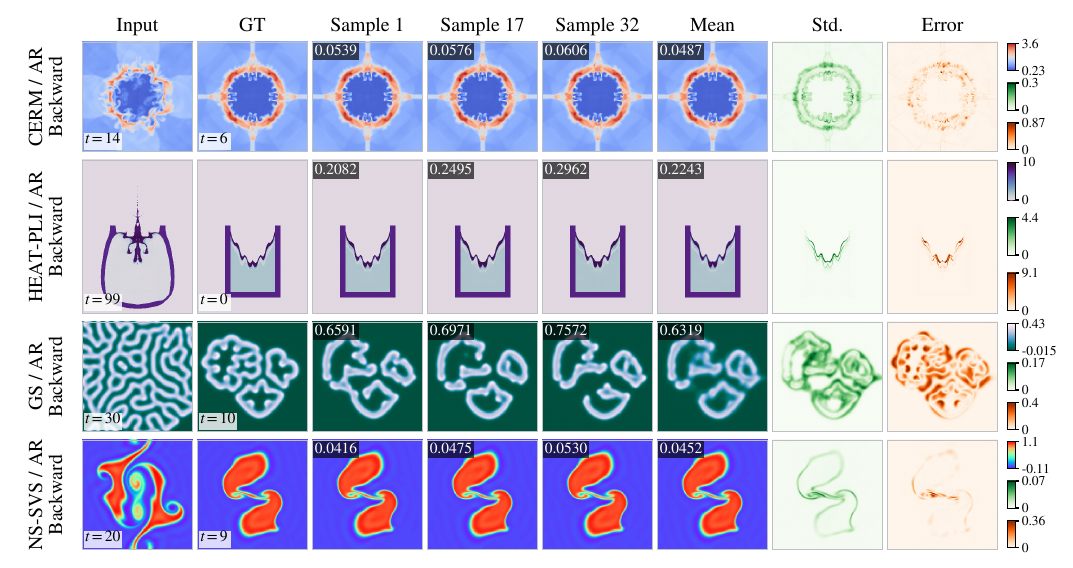}
    \caption{
    Forward and backward ensemble predictions of DiffUNet$^2$-AR on four datasets. Forward prediction is conditioned on the earlier frame and predicts the future frame, while backward prediction is conditioned on the last frame and predicts an earlier state.
    Sample 1, 17, and 32 denote members ranked by nRMSE, representing the best, middle-ranked,
    and worst predictions, respectively.
    Mean and standard deviation (Std.) are computed from all 32 samples.
    Error shows the absolute error of the ensemble mean relative to the
    ground truth. White numbers report nRMSE.
    }
    \label{fig:ensemble}
\end{figure*}

\subsection{Probabilistic Ensemble Quality}

We next evaluate the probabilistic prediction quality of DiffUNet$^2$ and the three probabilistic baselines.
\autoref{tab:distribution_quality} reports the results using 32 samples that are randomly generated from different noise initializations for each query.
CRPS measures the overall quality of the predicted distribution by comparing the generated ensemble with the ground truth. SSR compares the ensemble spread with the error of the ensemble mean, where values closer to one indicate better agreement between the two.
We further report the Spearman correlation $\rho$ between the spatial ensemble standard deviation and the absolute error of the ensemble mean.
This measures whether regions with larger variation across generated samples also tend to contain larger prediction errors.
For forward prediction, DiffUNet$^2$-AR achieves the lowest CRPS on all four datasets, reducing CRPS by 20.7--50.0\% relative to the best performance baseline on each dataset.
For backward prediction, DiffUNet$^2$-DR performs best on CERM, HEAT-PLI,
and Gray--Scott, while DiffUNet$^2$-AR performs best on NS-SVS, with
reductions of 20.9--48.6\% over the corresponding best baselines. his result follows the same pattern as the prediction-accuracy evaluation. SSR results are more dependent on the dataset and prediction direction.
DiffUNet$^2$ shows well-matched spread in several settings, such as forward CERM, while lower ratios occur on HEAT-PLI and several backward tasks. For $\rho$, positive value are observed across all evaluated settings, showing that regions with larger sample variation often also contain larger prediction errors and the strength varies across datasets and models.

\autoref{fig:ensemble} further shows how this variation appears in
the generated fields. Across the four datasets, the ensemble members preserve the main spatial structures while differing in local configurations. On HEAT-PLI, the outer material shape remains relatively stable, while larger variation appears around fine interior structures and local interfaces. Gray--Scott shows greater variation in the detailed reaction--diffusion patterns, whereas CERM and NS-SVS show differences mainly around evolving
interfaces and local flow structures. The standard-deviation maps show these regions of ensemble variation, and their overlap with the error maps is consistent with the positive spatial correlations reported in \autoref{tab:distribution_quality}. These examples show that DiffUNet$^2$ represents multiple possible spatial structures through stochastic generation, which has also been used for predictive uncertainty modeling in diffusion-based ensemble forecasting~\cite{li2024SEEDS}. Regions with larger ensemble variation often overlap with areas of higher prediction error.
Overall, DiffUNet$^2$ achieves consistently strong probabilistic prediction
quality in terms of CRPS in both temporal directions, while ensemble spread
and spatial variation show more dataset-dependent behavior.

\subsection{Extensions to Volumetric Data}

\begin{figure}[t]
    \centering
    \includegraphics[width=\columnwidth]{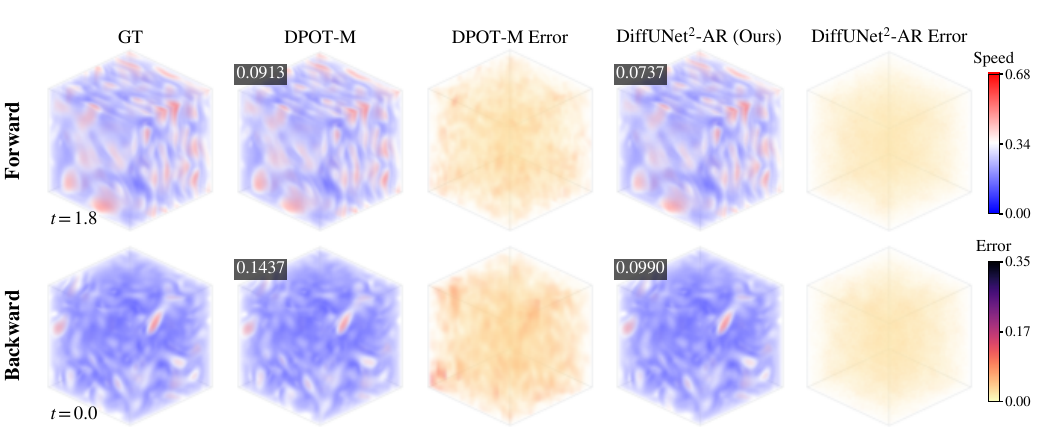}
    \caption{
    Volumetric prediction on 3D Kolmogorov Flow.
%The top and bottom rows show forward and backward prediction. White numbers report nRMSE.    
The top and bottom rows show forward and backward prediction, respectively.
Volumes visualize velocity magnitude and corresponding absolute error.
White numbers report full-volume nRMSE.
}
    \label{fig:3d_prediction}
\end{figure}

To broaden the applicability of DiffUNet$^2$ to volumetric scientific data, we further evaluate a native 3D implementation on Kolmogorov Flow. Since few of baselines provide native 3D support, we only compare with DPOT-M, which has a released 3D implementation.
As shown in Table~\ref{tab:3d_comparison}, DiffUNet$^2$-AR achieves the best
results across nRMSE, SSIM, and PSNR in both temporal directions.
Its forward and backward nRMSE values are 0.0607 and 0.0630, compared with
0.0709 and 0.0910 for DPOT-M.
DiffUNet$^2$-DR is less accurate in nRMSE, although it remains competitive in
SSIM and PSNR. Figure~\ref{fig:3d_prediction} further shows that DiffUNet$^2$-AR can preserve accurate volumetric flow structures in both forward and backward prediction.

% Requires \usepackage{booktabs}
\begin{table}[t]
\centering
\caption{
Prediction accuracy on Kolmogorov flow at $64^3$ resolution.
Best results are shown in bold.
}
\label{tab:3d_comparison}
\fontsize{5}{5}\selectfont
\setlength{\tabcolsep}{1.5pt}
\renewcommand{\arraystretch}{0.95}
\setlength{\aboverulesep}{1pt}
\setlength{\belowrulesep}{1pt}

\begin{tabular*}{\columnwidth}
{@{\extracolsep{\fill}}lcccccc@{}}
\toprule
\textbf{Method}
& \multicolumn{3}{c}{\textbf{Forward}}
& \multicolumn{3}{c}{\textbf{Backward}} \\
\cmidrule(lr){2-4}\cmidrule(lr){5-7}
& nRMSE $\downarrow$ & SSIM $\uparrow$ & PSNR $\uparrow$
& nRMSE $\downarrow$ & SSIM $\uparrow$ & PSNR $\uparrow$ \\
\midrule
DPOT-M
& 0.0709 & 0.9804 & 45.3094
& 0.0910 & 0.9706 & 42.8492 \\
DiffUNet$^2$-DR (Ours)
& 0.1141 & 0.9852 & 45.2545
& 0.1112 & 0.9781 & 45.7209 \\
DiffUNet$^2$-AR (Ours)
& \textbf{0.0607} & \textbf{0.9914} & \textbf{49.9176}
& \textbf{0.0630} & \textbf{0.9873} & \textbf{49.5818} \\
\bottomrule
\end{tabular*}
\end{table}

\subsection{Ablation Study}

\input{table/table3}

We evaluate the effects of joint bidirectional supervision and ambiguity-preserving cycle consistency on four datasets.
For both DiffUNet$^2$-DR and DiffUNet$^2$-AR, we compare three training
settings. \textit{Separate training} uses independent models for the two temporal directions. \textit{Joint training} shares one model across both directions. \textit{Cycle training} adds ambiguity-preserving cycle training. We compare these settings using prediction error and probabilistic ensemble metrics (\autoref{tab:ablation}).

We first compare separate and joint training.
Joint training uses one shared model for both temporal directions instead of training forward and backward models separately. The effect varies across datasets, but the shared model maintains comparable prediction accuracy.
On CERM, for DiffUNet$^2$-DR, bidirectional joint training reduces backward nRMSE by 22.2\%, from 0.05 to 0.0389. For DiffUNet$^2$-AR, forward and backward nRMSE are reduced by 11.1\%
and 13.8\%. Joint AR training also improves backward prediction on NS-SVS by 9\%.
On HEAT-PLI and Gray--Scott, the changes are small in both directions.
The probabilistic ensemble metrics show a similar trend.
On CERM, joint training reduces CRPS in both temporal directions for both
variants, with reductions of 5.4\% and 9.8\%. On NS-SVS, joint AR training further reduces backward CRPS by 10.0\%. On HEAT-PLI, joint and separate training perform similarly across both prediction and ensemble metrics. Gray--Scott shows a mixed effect, with some directions improving and others degrading. 
We next evaluate the effect of cycle consistency on the jointly trained models. For DiffUNet$^2$-DR  backward prediction, adding cycle training reduces backward nRMSE by 6.5\% on CERM, 6.7\% on
Gray--Scott, and 5.7\% on HEAT-PLI, with a smaller improvement on NS-SVS.
Backward CRPS also decreases on all four datasets, including reductions of
11.2\% on Gray--Scott and 9.4\% on HEAT-PLI.
For DiffUNet$^2$-AR, the changes are smaller overall and vary across datasets.
These results show that cycle consistency mainly benefits direct backward prediction.
%The ablation results confirm the value of the two main trainingcomponents in DiffUNet$^2$.
%The results support the effectiveness of the two main
%components in DiffUNet$^2$.
%Joint bidirectional supervision enables forward and backward prediction within
%a single shared model and provides gains on several prediction tasks.
%Although the improvement varies across datasets, the bidirectional setting maintains competitive performance without requiring separate directional models. The ambiguity-preserving cycle objective further show benefits to DR backward prediction, improving both nRMSE and CRPS across all four datasets.

\section{Application and Feedback}

This section shows the practical value of DiffUNet$^2$ through its application to a real scientific analysis workflow.
We collaborated with a scientific team studying deformation in multi-material systems and applied DiffUNet$^2$ to their analysis of HEAT-PLI simulations.

An important analysis task in their workflow is to search for an initial material-interface
geometry that produces a desired morphology at a later stage.
Different initial interface shapes can lead to different material deformation after shock loading.
The material morphologies of interest are often not available in the
existing simulation data.
Finding a desired outcome typically requires repeatedly modifying the initial geometry, running a high-fidelity simulation, and checking the resulting material state.
Because the simulations are expensive and the available trajectories cover
only a limited set of initial configurations, this process can require many
trial-and-error runs.

DiffUNet$^2$ bidirectional ability and target guided generation ability can accelarate this process. For states already available from simulation, the model acts as a surrogate for predicting their forward and backward temporal evolution.
For target states that are not present in the existing data, scientists can directly edit a material state toward a morphology of interest and refine the edited result through diffusion to obtain a physically plausible state under the learned data distribution (\autoref{fig:refin}).
The user-edited state can then be used as a new temporal condition, allowing the
model to infer plausible earlier configurations or explore its subsequent
evolution. 
Instead of searching only from initial conditions through repeated simulation,
the workflow allows scientists to reason backward from the outcome of interest
and narrow the configurations that require further high-fidelity simulation.
The probabilistic generation is also useful for examining fine-scale material
structures when the available data do not fully determine the underlying dynamics.
While the outlier material morphology can remain similar, interior
local structures may have several plausible configurations.
DiffUNet$^2$ exposes these alternatives through multiple generated samples,
allowing scientists to compare possible fine-scale morphologies and follow each
candidate through different temporal branches.

Two scientists from the collaborating team used the workflow in the context of
their HEAT-PLI analysis and collected feedback from other members of their
research group. They identified backward inference as particularly useful because their
analysis often starts from a desired final morphology rather than a known
initial configuration.
The generated candidate states provide possible starting configurations that
can be examined before running additional high-fidelity simulations.
They also found probabilistic generation useful for fine-scale material
structures.
Because several local structures can be plausible under the available
observations, viewing multiple generated samples helps them examine alternative
material configurations instead of relying on a single predicted state.
The scientists also valued the branching exploration provided by the interface (\autoref{fig:interface}). Different backward-generated candidates can be followed forward again, making it possible to compare how alternative initial configurations may evolve toward similar or different material morphologies. Together, target-guided editing, backward inference, and branching prediction provide an exploratory step before expensive simulation (\autoref{fig:teaser}). 
In this workflow, DiffUNet$^2$ provides scientists with an exploratory step before expensive simulation and reduces repeated trial-and-error when searching for a desired material outcome.

% remove this section, user study means too much extra
%\section{User Feedback}

%We conducted a user study to evaluate whether our model and the interactive system are useful for practical scientific analysis. We recruited [TO FILL: number] participants, including [TO FILL]. After a short tutorial, participants used the system on [TO FILL: datasets, e.g., HEAT-PLI, Wildfire, RealPDE] to complete three representative tasks: identifying deterministic and ambiguous regions from generated ensembles, performing what-if analysis through bidirectional generation and state editing, and exploring multiple plausible outcomes in probability space. During the study, we collected [TO FILL], followed by a short interview about the usefulness of the system and possible improvements.
%[To Do]

\section{Discussion and Future Work}

Our results show that bidirectional scientific temporal prediction can be
modeled within a shared probabilistic framework without sacrificing prediction
accuracy.
DiffUNet$^2$ performs strongly in both temporal directions and produces
high-quality probabilistic ensembles, while the AR and DR variants show
different advantages across datasets and prediction directions.
The ambiguity-preserving cycle objective provides its clearest improvement for
direct temporal backward prediction.
The practical extensions further adapt DiffUNet$^2$ for target-guided state generation, different scientific data representations, and interactive bidirectional exploration, allowing the model to support scientific analysis beyond prediction from fixed observations.
The current model remains data-driven, and its performance depends on the
coverage of the training distribution.
Ensemble spread and its relation to prediction error also vary across datasets,
and broader evaluation on volumetric data is still needed.
Future work will explore stronger physical constraints, improved probabilistic
calibration, and larger-scale scientific datasets.

\section{Conclusion}

We presented DiffUNet$^2$, a bidirectional conditional diffusion framework for
probabilistic scientific temporal modeling.
Experiments across multiple scientific datasets show strong forward and
backward prediction accuracy and high-quality probabilistic predictions within
a single shared model.
By modeling temporal relations in both directions, DiffUNet$^2$ supports both
prediction of future evolution and inference of plausible prior states.
We further extend the framework to target-guided state generation and interactive
temporal exploration.
These results demonstrate the potential of bidirectional generative modeling
as a flexible tool for scientific prediction and exploration.

\begin{comment}

\begin{figure}[h]
    \centering
    \includegraphics[width=0.9\linewidth]{figs/todo2.png}
    \caption{todo}
    \label{fig:example}
\end{figure}
\end{comment}

%\section{Discussions}

%% if specified like this the section will be omitted in review mode
%\acknowledgments{%
%	The authors wish to thank A, B, and C.
%  This work was supported in part by a grant from XYZ (\# 12345-67890).%
%}

\bibliographystyle{abbrv-doi-hyperref}

\bibliography{template}

%\clearpage
\appendix 

\makeatletter
\setlength{\@fptop}{0pt}
\setlength{\@fpsep}{8pt}
\setlength{\@fpbot}{0pt plus 1fil}
\setlength{\@dblfptop}{0pt}
\setlength{\@dblfpsep}{8pt}
\setlength{\@dblfpbot}{0pt plus 1fil}
\makeatother

\crefalias{section}{appendix} 
\onecolumn
\section{Appendix}

\section{Additional Results}
\label{app:additional_results}

\begin{figure*}[htbp]
    \centering
    \includegraphics[width=\textwidth]{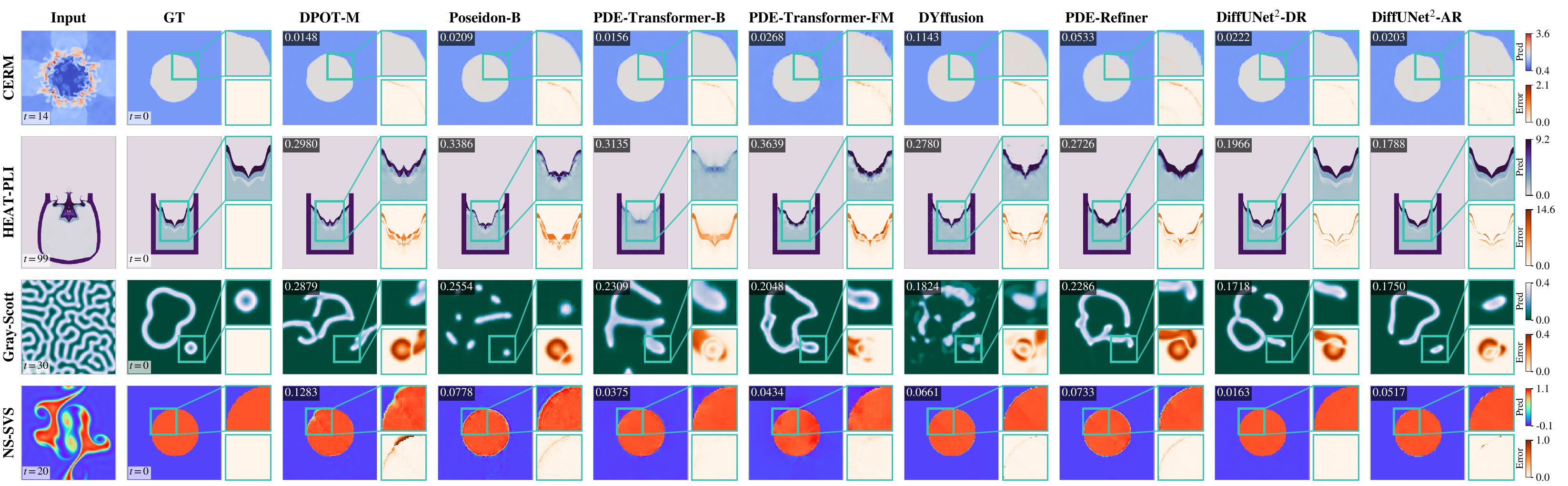}
    \caption{Backward prediction on four 2D datasets, conditioned on
    the last frame to infer the first frame. The outlined region is enlarged beside each field, with its absolute-error map below.}
    \label{fig:appendix_backward_endpoints}
\end{figure*}
\begin{figure*}[htbp]
    \centering
    \includegraphics[width=\textwidth]{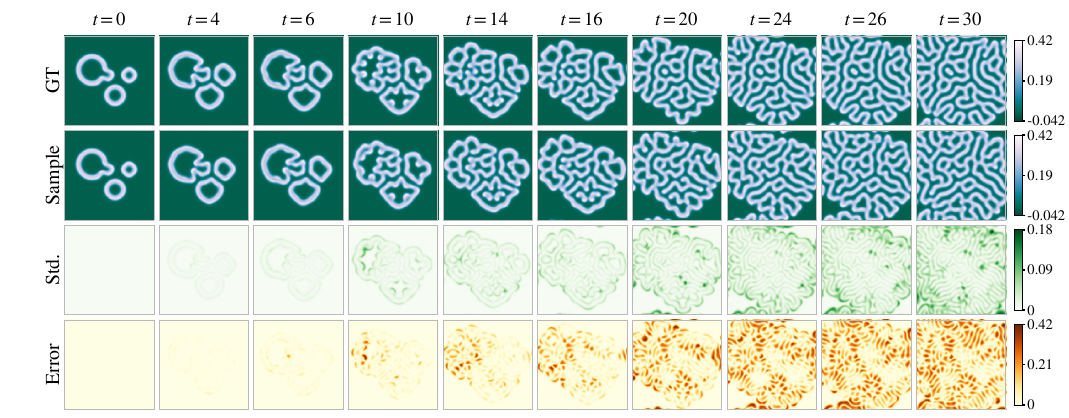}
    \caption{Forward prediction of Gray--Scott concentration $v$ with DiffUNet$^2$-DR, conditioned on the first frame. Rows show ground truth, ensemble member 17, standard deviation over all 32 members, and the absolute error of the displayed member.}
    \label{fig:appendix_gray_scott_dr_sequence}
\end{figure*}
\begin{figure*}[htbp]
    \centering
    \includegraphics[width=0.75\textwidth]{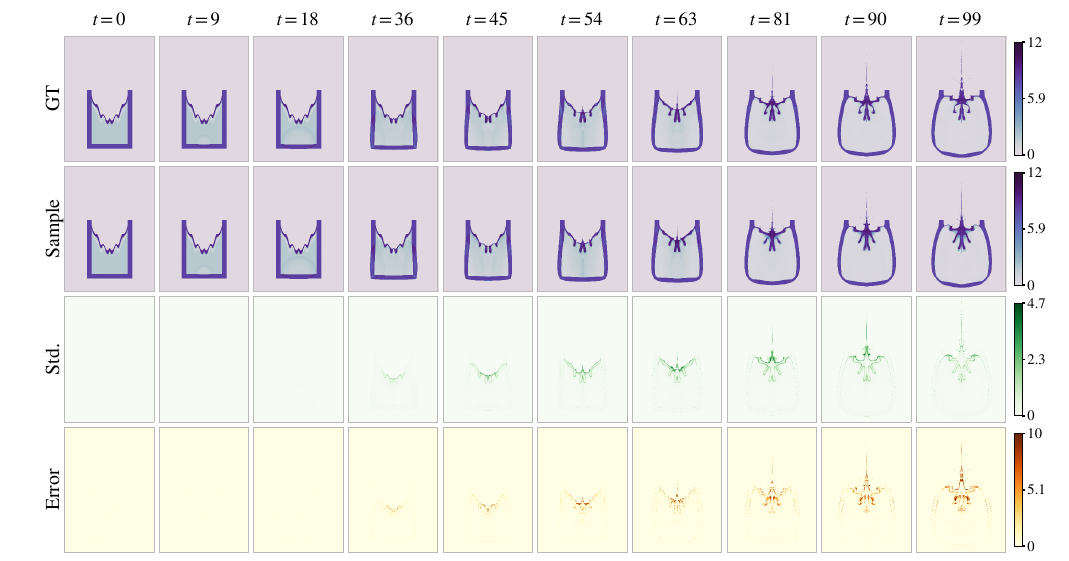}
    \caption{Forward prediction of HEAT-PLI density with DiffUNet$^2$-DR, conditioned on the first frame. Rows show ground truth, ensemble member 17, standard deviation over all 32 members, and the absolute error of the displayed member.}
    \label{fig:appendix_heat_dr_sequence}
\end{figure*}

\begin{figure*}[htbp]
    \centering
    \includegraphics[width=0.75\textwidth]{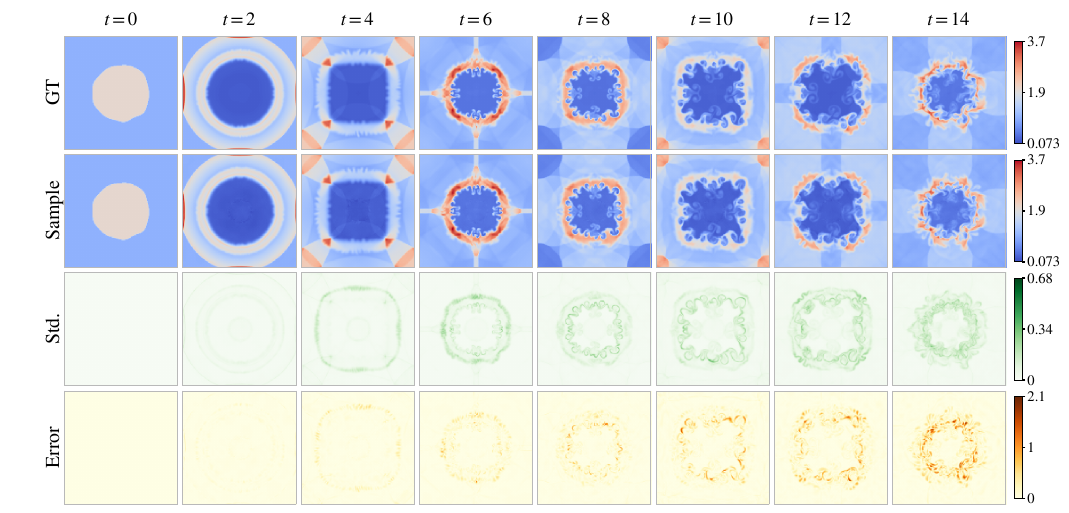}
    \caption{Forward prediction of CERM density with DiffUNet$^2$-DR, conditioned on the first frame. Rows show ground truth, ensemble member 17, standard deviation over all 32 members, and the absolute error of the displayed member.}
    \label{fig:appendix_cerm_dr_sequence}
\end{figure*}

\begin{figure*}[htbp]
    \centering
    \includegraphics[width=0.9\textwidth]{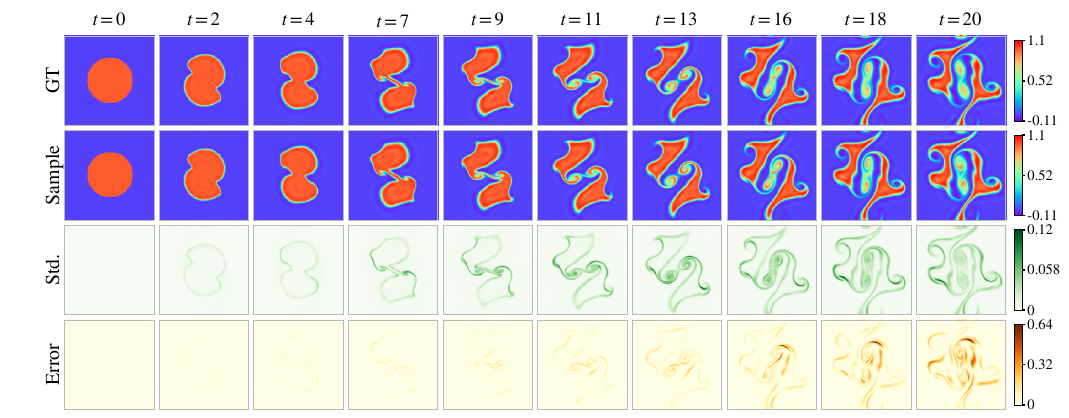}
    \caption{Forward prediction of the NS-SVS passive tracer with DiffUNet$^2$-DR, conditioned on the first frame. Rows show ground truth, ensemble member 17, standard deviation over all 32 members, and the absolute error of the displayed member.}
    \label{fig:appendix_ns_svs_dr_sequence}
\end{figure*}

\begin{figure*}[htbp]
    \centering
    \includegraphics[width=\textwidth]{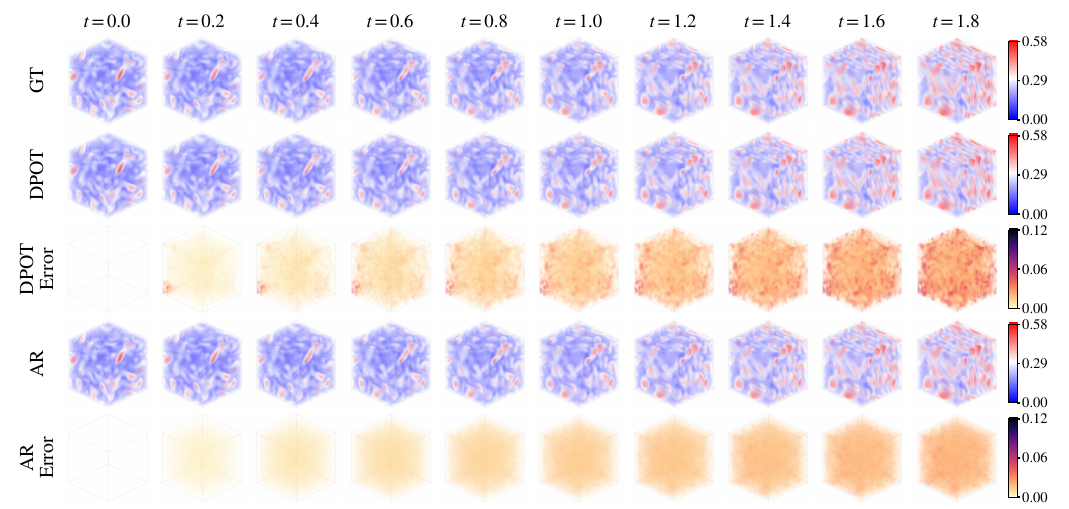}
    \caption{Forward prediction of 3D Kolmogorov flow, conditioned on the initial state. Rows show ground truth, DPOT-M prediction and error, and DiffUNet$^2$-AR prediction and error. State volumes show speed; AR uses the mean velocity over 16 samples. Error volumes show the magnitude of the velocity-vector error. Time increases from left to right. Both models share the same camera, state color scale, and error color scale across all time points.}
    \label{fig:appendix_kolmogorov_ar_sequence}
\end{figure*}

\begin{figure*}[htbp]
    \centering
    \includegraphics[width=\textwidth]{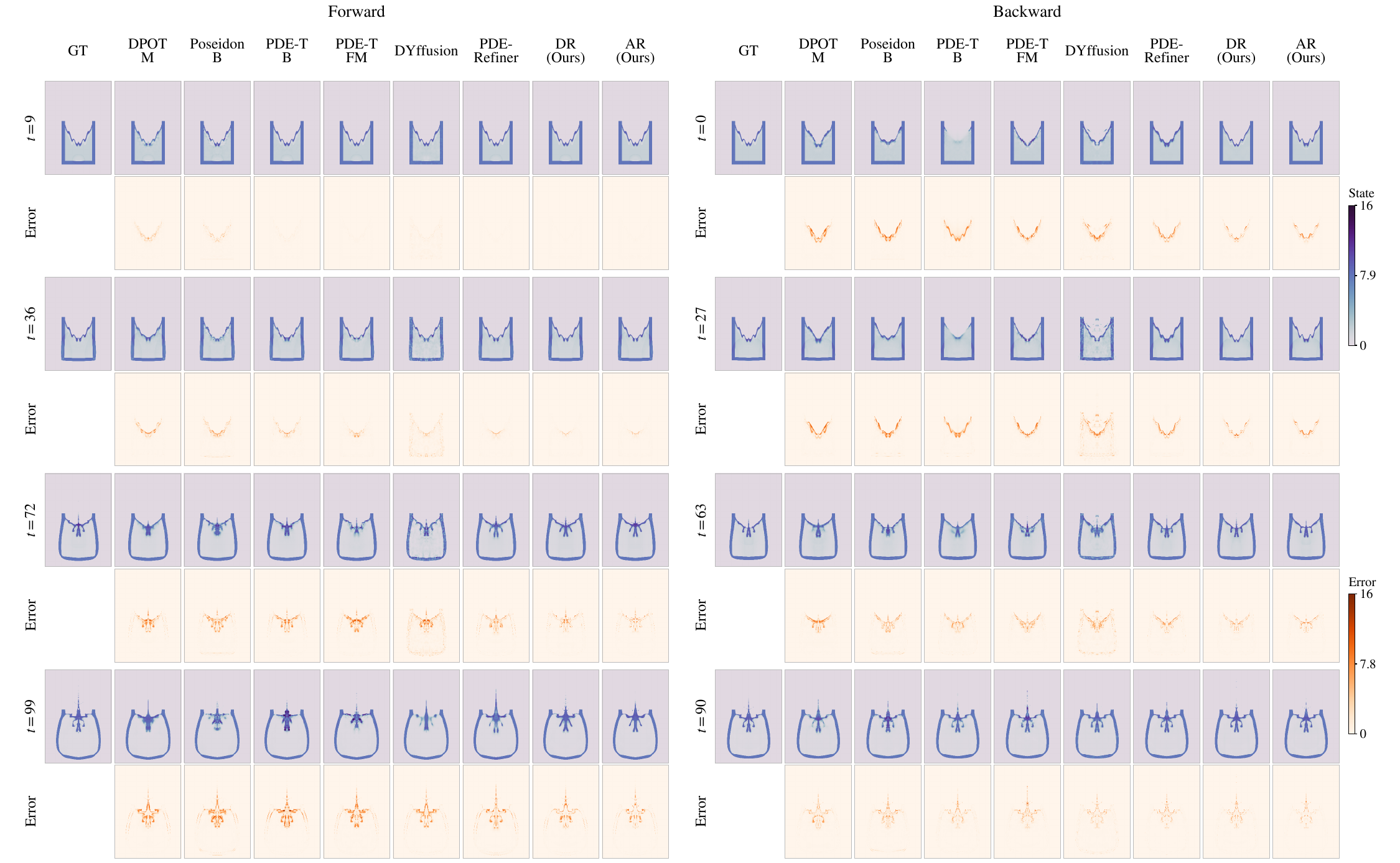}
    \caption{Forward (left) and backward (right) prediction of HEAT-PLI density, conditioned on the first and last frames, respectively. Four queried times increase from top to bottom. At each time, ground truth and eight model predictions appear above the corresponding absolute-error maps. Probabilistic models show ensemble member 17 of 32. PDE-T denotes PDE-Transformer; DR and AR denote the two DiffUNet$^2$ variants. State and error maps each share a color scale across models, directions, and times.}
    \label{fig:appendix_heat_model_grid}
\end{figure*}

\begin{figure*}[htbp]
    \centering
    \includegraphics[width=\textwidth]{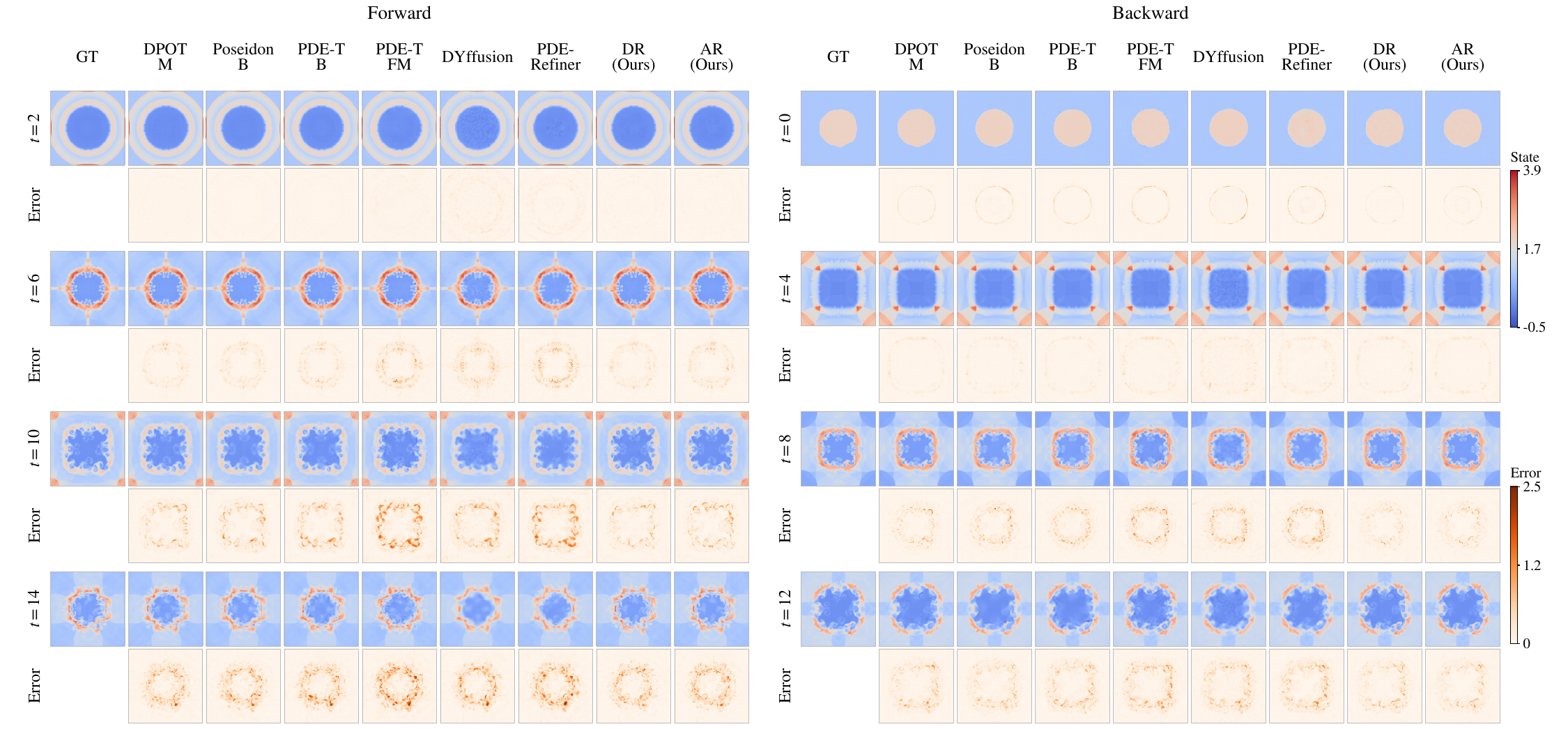}
    \caption{Forward (left) and backward (right) prediction of CERM density from the first and last frames, respectively. Each time point occupies two rows: ground truth and predictions above, absolute errors below. Time increases downward in both blocks. Probabilistic predictions use ensemble member 17 of 32. PDE-T denotes PDE-Transformer, and DR and AR denote DiffUNet$^2$. State and error maps use separate scales, each fixed across all models, directions, and times.}
    \label{fig:appendix_cerm_model_grid}
\end{figure*}

\begin{figure*}[htbp]
    \centering
    \includegraphics[width=\textwidth]{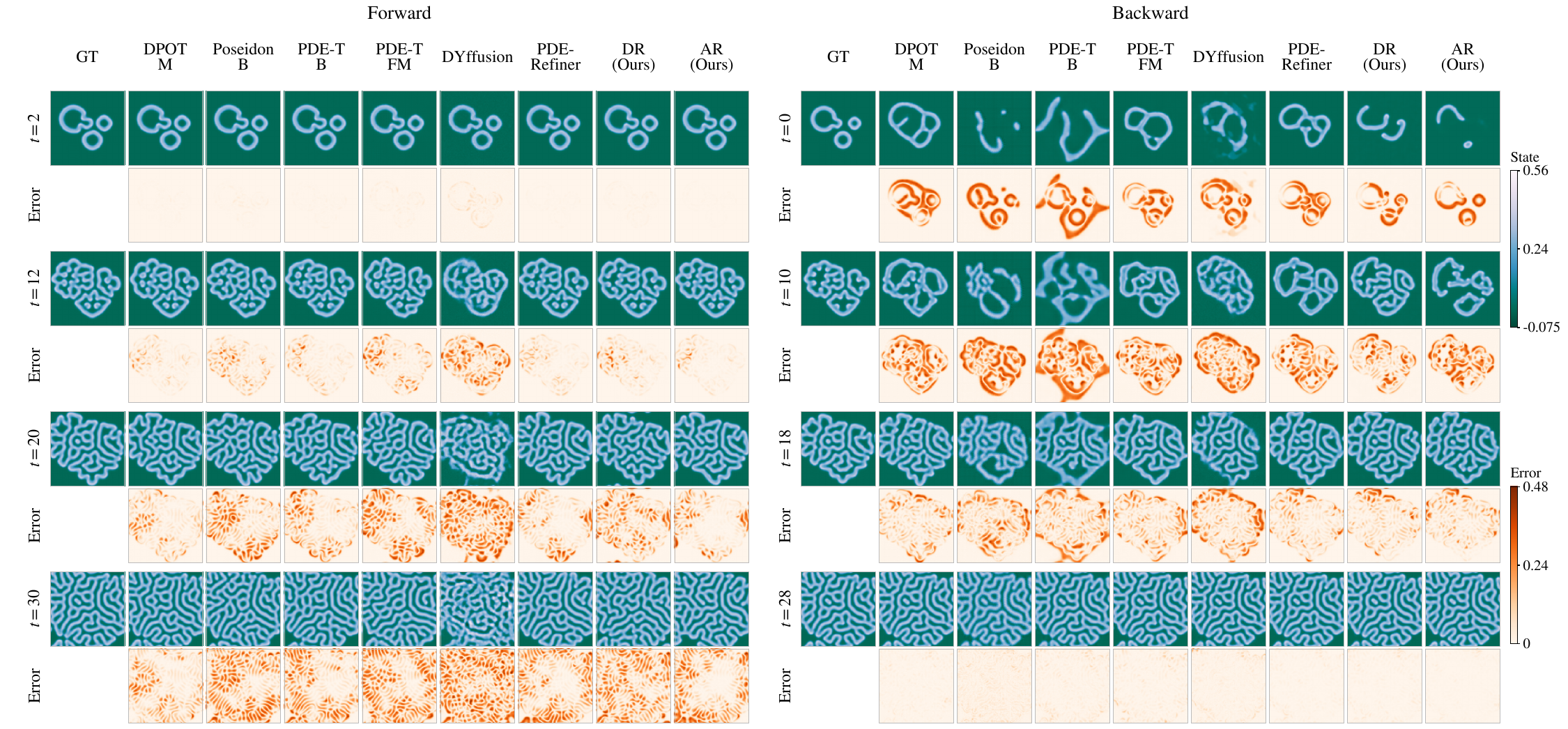}
    \caption{Forward (left) and backward (right) prediction of Gray--Scott concentration $v$, conditioned on the first and last frames, respectively. Four queried times increase from top to bottom, with each prediction row followed by its absolute-error maps. Columns show ground truth and eight models; probabilistic predictions use ensemble member 17 of 32. PDE-T denotes PDE-Transformer, and DR and AR denote DiffUNet$^2$. State and error scales are shared across models, directions, and times.}
    \label{fig:appendix_gray_scott_model_grid}
\end{figure*}

\begin{figure*}[htbp]
    \centering
    \includegraphics[width=\textwidth]{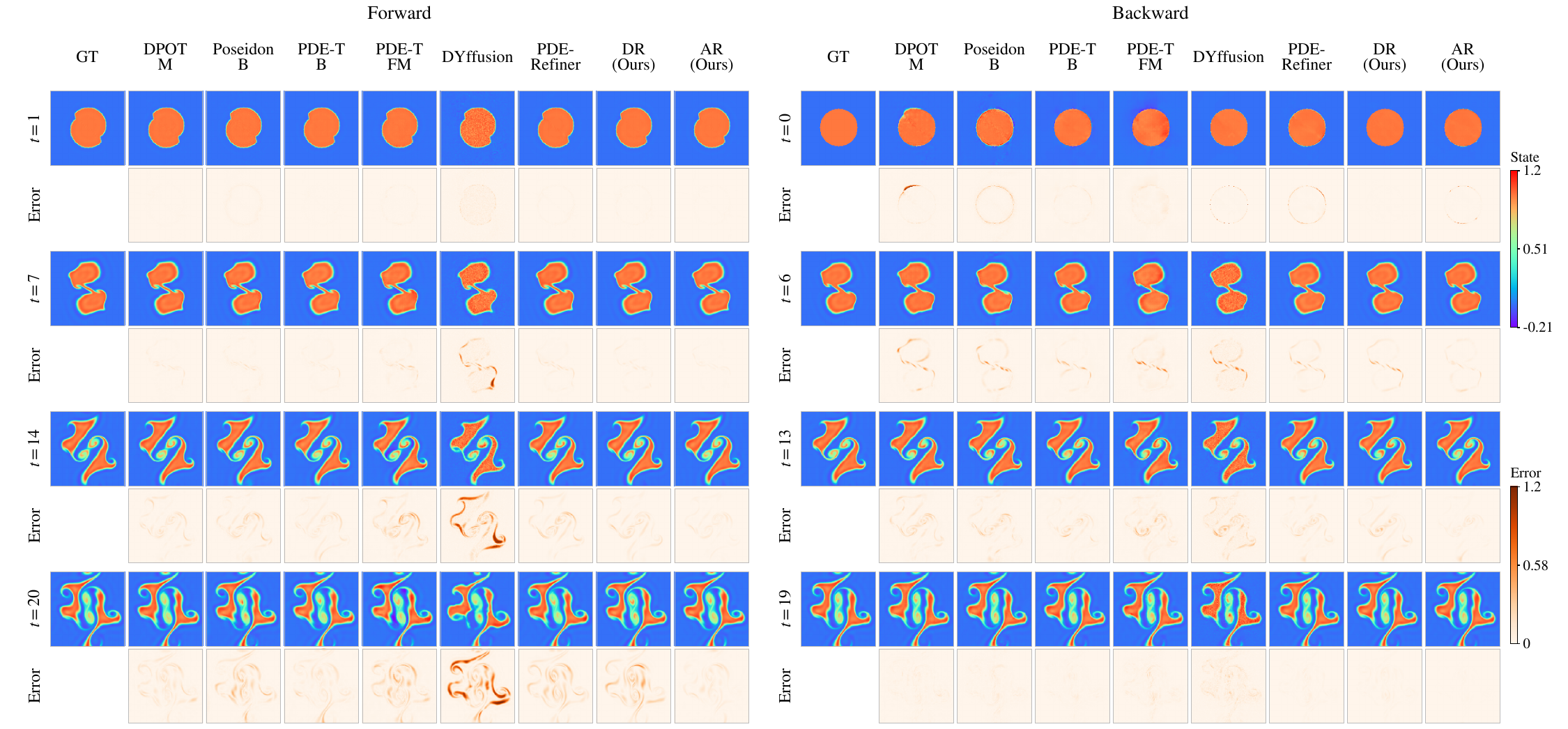}
    \caption{Forward (left) and backward (right) prediction of the NS-SVS passive tracer from the first and last frames, respectively. Time increases downward in both blocks. Each time point shows ground truth and eight predictions, followed by the absolute-error maps. Probabilistic predictions use ensemble member 17 of 32. PDE-T denotes PDE-Transformer, and DR and AR denote DiffUNet$^2$. State and error maps each use one color scale across all models, directions, and times.}
    \label{fig:appendix_ns_svs_model_grid}
\end{figure*}

% Additional visualization slot 2.
% \begin{figure*}[htbp]
%     \centering
%     \includegraphics[width=\textwidth]{appendix_figures/additional_results_02.pdf}
%     \caption{...}
%     \label{fig:appendix_additional_02}
% \end{figure*}

\end{document}

%% file: content/1_introduction.tex
Analyzing the temporal evolution of physical systems, ranging from hydrodynamics and thermal transport to geophysical dynamics, is important for scientific discovery~\cite{raissi2019physics, casiraghi2024complexsystem}. 
Traditionally, researchers rely on simulations or experiments, with visualization used
to examine the resulting phenomena and guide iterative hypothesis refinement ~\cite{moreland2013pipe}.
However, this process can be limited by the computational cost of simulations and the expense or limited availability of experiment data and observations~\cite{schneider2017climate}.
Machine learning models are therefore increasingly used as surrogate models~\cite{Anirudh2020,yitang2025INR}, approximating system behavior to reduce the cost of repeated simulation and accelerate scientific analysis.

While machine learning surrogates effectively improve efficiency, gaps remain between current temporal modeling methods and practical scientific analysis workflows. 
Most existing temporal surrogate models are designed for deterministic forward prediction, producing a single future state from a given condition.
However, the temporal evolution of complex systems is often difficult to predict deterministically due to limitations in observation or modeling. Unknown governing mechanisms, incomplete observations, measurement noise, and chaotic or stochastic dynamics can all lead to multiple plausible states or divergent trajectories~\cite{strogatz2018nonlinear}. When multiple outcomes are plausible, a single deterministic prediction cannot represent the range of possible states and may average over distinct evolution
patterns.
In such cases, temporal prediction need represent a distribution over plausible outcomes rather than a single point estimate.
Furthermore, most temporal models are designed only for forward prediction, while scientific analysis often requires reasoning in both temporal directions. 
For example, a later observation may be used to infer plausible prior states
and trace how the observed outcome developed.
Backward inference is often more difficult because information can be lost during physical evolution. Processes such as diffusion, dissipation, and mixing can make different prior states evolve into similar later observations, leading to a non-unique inverse problem~\cite{stuart2010inverse,min2014backwardheat}.
These gaps motivate a unified temporal surrogate model that supports probabilistic prediction in both forward and backward directions.

Recently, diffusion models have demonstrated a strong ability to model complex data distributions and generate diverse, high-fidelity samples~\cite{ho2020ddpm, song2021score, nichol2021iddpm, song2022ddim}. 
By denoising from different noise initializations, diffusion models can generate multiple plausible outcomes from the learned distribution rather than reducing them to a single average prediction.
This property has motivated their use in scientific domains, especially for data with ambiguity or uncertainty~\cite{li2024SEEDS, cachay2023dyffusion,yu2026probabilistic, andrae2025continuous}. 
In such settings, diffusion models can capture plausible evolutions that are often hidden by deterministic averaging, providing a new lens for scientific exploration.
This property has motivated their use in scientific temporal modeling~\cite{li2024SEEDS,shu2024zeroshot}. When the evolution is well constrained, different samples can remain similar, when multiple outcomes are possible, the samples can reflect different
plausible states.
However, existing approaches are mainly developed for forward temproal prediction ~\cite{cachay2023dyffusion,yu2026probabilistic, andrae2025continuous}. 
Less attention has been given to modeling both temporal relations within a shared model. The backward direction further requires the model to preserve multiple possible prior states while keeping them consistent with the later observation.

To support probabilistic prediction in both temporal directions, we introduce DiffUNet$^2$ (\autoref{fig:model}), a bidirectional conditional diffusion surrogate model for scientific temporal data.
Given an observed state and a target time, DiffUNet$^2$ models the conditional distribution of the queried state in either temporal direction. A shared residual diffusion model is trained on both forward and backward queries.
We consider two common temporal prediction settings through bidirectional supervision for autoregressive rollout and direct temporal prediction. To account for the non-uniqueness of backward inference, we further introduce an ambiguity-preserving cycle consistency supervision that maintains consistency between temporal directions without requiring a backward prediction to recover a single recorded history.
We evaluate DiffUNet$^2$ on four scientific temporal datasets covering compressible and incompressible fluid dynamics
(CERM and NS-SVS~\cite{herde2024poseidon}), chemical reaction--diffusion (Gray--Scott~\cite{pearson1993patterns}),
and shock-driven material deformation (HEAT-PLI~\cite{Banesh2025HEATPLI}).
We compare DiffUNet$^2$ with six representative temporal baselines, including three deterministic and three probabilistic models~\cite{hao2024dpot,herde2024poseidon,holzschuh2025pdetransformer,lipman2023flowmatching,lippe2023pderefiner,cachay2023dyffusion}.
Results show that DiffUNet$^2$ achieves strong predictive performance in both temporal directions and produces high-quality probabilistic ensembles across the evaluated datasets. 
Beyond temporal prediction, we extend DiffUNet$^2$ with target-guided state editing for practical scientific exploration.
A scientific state can be modified to express a desired feature, refined under the learned data distribution, and then used to explore possible prior and future states.
We provide a interface for accessing the model capabilities, and case studies with domain scientists shows how bidirectional prediction and state editing can support scientific temporal analysis. In summary,we have the following contributions:

\begin{itemize}
    \item We propose DiffUNet$^2$, a unified probabilistic framework for
    bidirectional scientific temporal modeling. A single conditional diffusion model learns both forward and backward temporal relations and supports autoregressive rollout and direct temporal prediction.

   % . DiffUNet$^2$ jointly models forward and backward evolutions within a shared model and supports both autoregressive rollout and direct temporal prediction.
    %\item We introduce DiffUNet$^2$, a unified probabilistic framework for
    %bidirectional scientific temporal modeling. A single conditional diffusion model learns both forward and backward temporal relations

    \item We introduce a bidirectional supervision strategy with ambiguity-preserving cycle consistency, enabling joint learning of forward and backward temporal relations while retaining multiple plausible outcomes.
\item We extend DiffUNet$^2$ from temporal prediction to practical scientific exploration by designing a target-guided state editing method and an interactive interface to make the model easier to use in real analysis workflows, allowing scientists to edit states of interest and explore their plausible prior and future evolution.
% \item We extend DiffUNet$^2$ from temporal prediction to practical scientific exploration. States of interest can be specified and refined under the learned data distribution, then reused for bidirectional temporal queries to explore possible prior and future evolution beyond recorded or simulated trajectories.

%\item \textit{A generative framework for accelerating scientific exploration}. We show how generated scientific states can be reused as conditions for further temporal queries. By iteratively generating, selecting, and extending states across time, the framework supports efficient exploration of alternative evolutions without requiring every trajectory to be obtained from simulations or experiments.

 %   \item We introduce a guided generation method that allows analysts to specify a phenomenon of interest and obtain plausible scientific states that reflect the desired outcome. Such states would otherwise require repeated simulation and manual parameter tuning to obtain. They can subsequently be used to infer possible histories or predict further evolution.

\end{itemize}

%% file: fig_model.tex
\begin{figure*}[h]
    \centering
    \includegraphics[width=\linewidth]{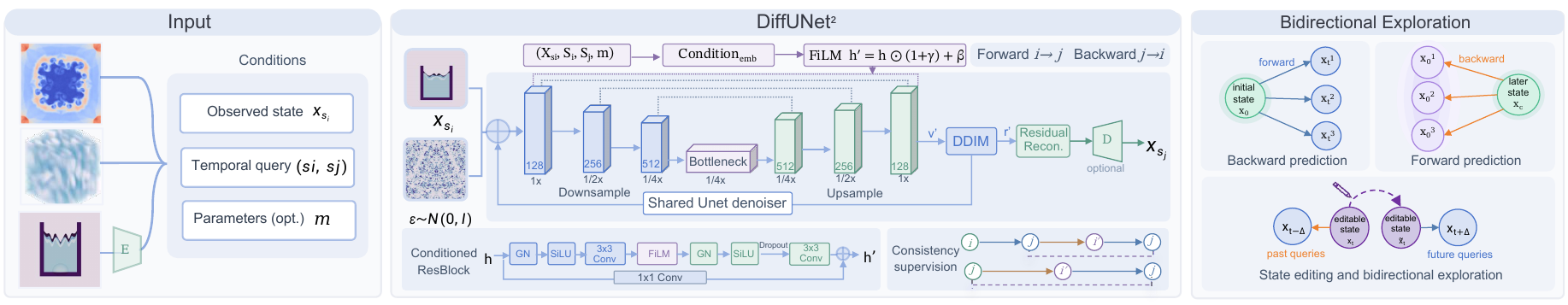}
    \caption{
    Overview of DiffUNet$^2$ for bidirectional temporal prediction and exploration. Given an observed state $X_{s_i}$, a temporal query $(s_i,s_j)$, and optional parameters $m$, a shared U-Net denoiser uses FiLM conditioning to predict the target state $X_{s_j}$ through DDIM sampling and residual reconstruction, with an optional decoder $D$. Consistency supervision supports prediction in both temporal directions. The resulting model enables backward and forward prediction, as well as interactive state editing and exploration of past and future states.
    %Input temporal data are optionally compressed by a KL-regularized VAE for efficient modeling. DiffUNet$^2$ then performs conditional diffusion generation under the observed state and temporal query, supporting bidirectional any-to-any prediction as well as user-guided editing and refinement through guided denoising. %The generated states are integrated into an interactive visual analysis system that supports branching temporal exploration, jump-time prediction, direct feature editing, and multi-branch probabilistic exploration.
    %From top to bottom, the rows show the \textit{ground-truth}, \textit{forward prediction} conditioned on the first frame, \textit{center-out prediction} conditioned on the middle frame, and \textit{backward prediction} conditioned on the last frame. The numbers beneath the predicted frames indicate the \textit{nRMSE} at that time step. The color bar represents the physical value in each heat map.
    }
    \label{fig:model}
\end{figure*}
%\end{sidewaysfigure}

%% file: content/2_related_works.tex
\section{Related Works}
%We frame the modeling of scientific dynamics as a learning-based surrogate problem in which the objective is to approximate the transition kernels of complex systems. 
%This section reviews the transition from deterministic simulators to probabilistic generative models and positions our work within the broader context of interactive ensemble visualization and data distribution space exploration.

\subsection {Scientific Temporal Data Modeling}
Learning-based surrogate models have become an efficient alternative to
model the scientific temporal evolution. 
Their applicability has expanded across diverse physical processes, such as PDE simulations, fluid dynamics, weather forecasting, and other scientific observation and simulation settings
~\cite{takamoto2022pdebench, li2021fourier, lam2023learning, liu2019solar}.
Benchmarks such as PDEBench~\cite{takamoto2022pdebench},
CFDBench~\cite{luo2023cfdbench}, and
APEBench~\cite{koehler2024apebench} provide diverse settings for evaluating
learned temporal surrogates across 
different dynamics, flow conditions, and spatial settings.
Recent work has focused on improving the generality of such models across
different physical systems.
DPOT~\cite{hao2024dpot} uses large-scale pretraining across multiple PDEs,
while Poseidon~\cite{herde2024poseidon} and
PROSE-PDE~\cite{sun2025prosepde} study transferable modeling across different
dynamics and operators.
PDE-Transformer~\cite{holzschuh2025pdetransformer} further develops scalable
temporal modeling across multiple physical fields.
These advances broaden the range of dynamics that can be modeled by
a learned surrogate and improve prediction across different systems and time
scales. Despite these advances, their method still primarily focuses on predicting future system evolution from earlier observations using a deterministic model.
%However, scientific analysis often involves reasoning about multiple possible trajectories arising from uncertainty in observations, parameters, or model assumptions, and infer the possible history states. In such scenarios, deterministic predictors provide limited support for exploring alternative hypotheses or understanding variability in temporal evolution.

\subsection{Inverse and Bidirectional Scientific Modeling}
Scientific analysis often requires reasoning in both temporal directions.
Forward prediction estimates how a system may evolve, while inverse reasoning
seeks the states, parameters, or conditions that could explain an observation~\cite{stuart2010inverse,raissi2019physics}.
Inverse problems are often ill-posed because the solution may be non-unique or
sensitive to small changes in the observations~\cite{stuart2010inverse}.
For example, in dissipative processes such as diffusion and mixing, different earlier states can evolve into similar later states as information is lost over time. Backward heat reconstruction is a classic example of this instability~\cite{min2014backwardheat}. As a result, backward prediction may need to represent multiple plausible prior states rather than a single deterministic solution.
Recent work has explored learning-based inverse and bidirectional modeling.
Physics-informed neural networks~\cite{raissi2019physics} and Neural Inverse
Operators~\cite{molinaro2023neuralinverse} infer unknown physical quantities
from observations, while invertible neural operators~\cite{long2025invertible}
couple forward and inverse mappings within a shared model.
Diffusion-based approaches further model distributions over possible inverse
solutions~\cite{haitsiukevich2024diffusion}.
Most existing methods, however, focus on parameter estimation and state recovery  while probabilistic backward prediction between
scientific states has received less attention.

%However, scientific analysis often involves reasoning about multiple possible trajectories arising from uncertainty in observations, parameters, or model assumptions. In such scenarios, deterministic predictors provide limited support for exploring alternative hypotheses or understanding variability in temporal evolution.

% For systems represented by particles or irregular meshes, graph-based formulations offer a natural abstraction for modeling local interactions. Sanchez-Gonzalez et al.~\cite{sanchez2020learning} demonstrated that Graph Network-based Simulators(GNS) can capture complex Lagrangian dynamics by learning message-passing updates over node-and-edge representations. Subsequent architectures, such as MeshGraphNets~\cite{pfaff2021learning}, extend this to multi-scale mesh structures to simulate aerodynamics and structural mechanics with high fidelity. By encoding inductive biases like permutation invariance and local connectivity, these models effectively generalize across different material properties and boundary conditions without requiring a regular grid. Despite their success, most learned simulators are designed for deterministic forward prediction. 

\subsection{Probabilistic Diffusion Models for Scientific Dynamics}
To address the limitations of deterministic models, earlier generative approaches such as Variational Autoencoders (VAEs)~\cite{kingma2019VAE} and Generative Adversarial Networks (GANs)~\cite{Goodfellow2014GAN} were explored to capture multi-modal outcomes.
Recent work has applied diffusion and related generative models to scientific
dynamics.
DYffusion~\cite{cachay2023dyffusion} combines temporal interpolation with
diffusion-based forecasting for stochastic dynamics, while
PDE-Refiner~\cite{lippe2023pderefiner} uses iterative denoising to refine PDE
predictions.
PDE-Transformer-FM~\cite{holzschuh2025pdetransformer,lipman2023flowmatching}
uses flow matching to model distributions over future physical states.
Generative ensemble methods have also been studied in weather forecasting,
such as SEEDS~\cite{li2024SEEDS}, where multiple samples represent possible
future evolutions.
In addition, diffusion models have also been shown to be naturally suited for uncertainty modeling~\cite{shu2024zeroshot}, as their stochastic sampling process supports confidence interval estimation and out-of-distribution detection without task-specific retraining. These studies show that diffusion models can represent alternative outcomes
and uncertainty through conditional sampling.
Compared with deterministic predictors, these models can preserve multiple plausible outcomes that follow the probabilistic distribution and reflect predictive uncertainty. However, most existing work still treats diffusion primarily as a predictive or sampling model. Less attention has been paid to how such probabilistic generation can be systematically incorporated into the practical workflows of scientific analysis, including ambiguity diagnosis, backward reasoning, and interactive hypothesis exploration.

%% file: content/3_background.tex
\section{Backgrounds and Model Design Rationale}
\label{sec:ChallengeGoals}

This section summarizes the modeling considerations that motivate DiffUNet$^2$. We first discuss why scientific temporal prediction can require probabilistic generation rather than deterministic prediction, and then consider the additional challenges introduced by bidirectional temporal inference and backward prediction.

\textbf{Probabilistic temporal modeling.}
Many learning-based temporal surrogates formulate temporal prediction as learning a deterministic solution operator that maps a given state or condition to a target future state~\cite{hao2024dpot, herde2024poseidon, koehler2024apebench}. However, in practical scientific settings, the target future state may not be uniquely determined by the available conditions, due to incomplete observations, unresolved variables, measurement limitations, or inherent stochasticity in the underlying dynamics. Therefore, we consider temporal prediction as modeling a conditional distribution over possible target states.

Diffusion models provide a generative formulation for this setting~\cite{cachay2023dyffusion, lippe2023pderefiner}. Conditioned on the available information, the denoising process starts from random Gaussian noise and progressively generates a sample from the learned data distribution.
Hence, different random noise initializations produce different plausible states under the same condition.
When the temporal relation is well constrained, different samples tend to converge to similar states. When the relation is under-determined, different samples can spread across multiple plausible outcomes.
The generated samples show both predictive uncertainty and the possible forms of the system state under the same condition.
 We use this probabilistic property as the basis of DiffUNet$^2$.

\textbf{Bidirectional Temporal Modeling}.
Most temporal prediction surrogate models are designed for forward prediction. However, scientific temporal analysis may require reasoning in both temporal directions, predicting future states from earlier observations and inferring possible prior states from later ones~\cite{min2014backwardheat, stuart2010inverse}. We seek a temporal model that supports both forward and backward queries within a unified formulation.
Forward and backward queries describe transitions over the same temporal process but in opposite directions. Sharing the model across both directions allows the temporal query to specify the source state, target time, and
direction without separate prediction models. This leads to the bidirectional temporal supervision described in Sec. \ref{Bidirectional Temporal Supervision}.

Furthermore, backward prediction introduces an inverse problem that can be more under-determined than forward prediction.
Many common physical processes, such as diffusion, dissipation, and mixing, progressively lost information over time. As a result, different previous states may produce similar later observations, making the inverse problem ill-posed.
This property also changes how consistency should be imposed between the two temporal directions. A backward-generated state should be consistent with the later observation when propagated forward. At the same time, this consistency should not enforce a unique backward solution, since multiple prior states may evolve into the same later observation~\cite{min2014backwardheat, stuart2010inverse}. This motivates the
ambiguity-preserving cycle objective introduced in Sec.\ref{Ambiguity-Preserving Cycle Consistency}.

\textbf{Practical scientific exploration.} Bidirectional temporal prediction allows users to reason from an available state, but researchers may be interested in a particular structure, morphology, or outcome that has not appeared in the existing data.
Obtaining such states can require repeated simulations and experiments. Diffusion models are well suited to this setting because they can model the distribution of plausible scientific states in addition to conditional temporal transitions. DiffUNet$^2$ allows direct modification of generated states to introduce desired structures or correct undesirable features. The edited states are then refined by DiffUNet$^2$ to suppress implausible artifacts while preserving the intended modifications, as described in Sec.\ref{Target-guided state generation and editing.}.

%% file: table/table_data.tex
\begin{table}[t]
\centering
\captionsetup{skip=3pt}
\caption{Evaluation datasets. Time spans use dataset's simulation units.}
\label{tab:eval-datasets}
\footnotesize
\setlength{\tabcolsep}{2pt}
\renewcommand{\arraystretch}{0.88}
\setlength{\aboverulesep}{1pt}
\setlength{\belowrulesep}{1pt}
% CERM's local time coordinate is zero-filled; use the source CE-RM
% schedule (frame 14 is t=1.4). Other spans follow stored simulation times
% or, for Kolmogorov Flow, the generator's initial state and 0.2 spacing.
% Data size = Traj. * T * Ch. * grid points * stored bytes per value / 2^30.
% HEAT uses float16; the remaining physical arrays use float32.
% Excludes unselected frames/channels, file headers, latent copies and predictions.
\resizebox{\columnwidth}{!}{\begin{tabular}{@{}lllrrlr@{}}
\toprule
Dataset & Fields (Ch.) & Grid & Traj. & $T$ & Time span & Data size (G) \\
\midrule
CERM~\cite{herde2024poseidon} & $\rho,u_x,u_y,p$ (4) & $128^2$ & 140 & 8 & $[0.00,1.40]$ & 0.273 \\
HEAT-PLI~\cite{Banesh2025HEATPLI} & Density (1) & $1120\times400$ & 220 & 12 & $[0.00,24.75]$ & 2.203 \\
Gray--Scott~\cite{pearson1993patterns} & $u,v$ (2) & $128^2$ & 140 & 16 & $[238.00,3298.00]$ & 0.273 \\
NS-SVS~\cite{herde2024poseidon} & $u_x,u_y,c$ (3) & $128^2$ & 140 & 21 & $[0.00,1.00]$ & 0.538 \\
Kolmogorov Flow~\cite{koehler2024apebench} & $u_x,u_y,u_z$ (3) & $64^3$ & 70 & 10 & $[0.00,1.80]$ & 2.051 \\
\bottomrule
\end{tabular}}
\end{table}

%% file: table/table1.tex
\definecolor{datasetshade}{RGB}{245,245,245}
\definecolor{rankteal}{RGB}{100,150,242}
\colorlet{rankfirst}{rankteal!30}
\colorlet{ranksecond}{rankteal!20}
\colorlet{rankthird}{rankteal!10}
\begin{table*}[t]
\centering
\caption{
Quantitative comparison on four datasets for forward and backward prediction.
Probabilistic models are evaluated using the mean of 32 samples, while
deterministic models use a single prediction.
Teal shading from darker to lighter indicates the top three results
for each dataset, metric, and prediction direction; only the best results are bold.
}
\label{tab:main_prediction_accuracy}
\fontsize{7}{7}\selectfont
% Retain column spacing so resizebox does not cancel the smaller type.
\setlength{\tabcolsep}{5.5pt}
\renewcommand{\arraystretch}{0.9}
\setlength{\aboverulesep}{1pt}
\setlength{\belowrulesep}{1pt}
\resizebox{\textwidth}{!}{
\begin{tabular}{lccc ccc ccc ccc}
\toprule
\textbf{Methods}
& \multicolumn{3}{c}{\textbf{CERM}}
& \multicolumn{3}{c}{\textbf{Heat}}
& \multicolumn{3}{c}{\textbf{Gray--Scott}}
& \multicolumn{3}{c}{\textbf{NS-SVS}} \\
\cmidrule(lr){2-4}
\cmidrule(lr){5-7}
\cmidrule(lr){8-10}
\cmidrule(lr){11-13}
& nRMSE $\downarrow$ & SSIM $\uparrow$ & PSNR $\uparrow$
& nRMSE $\downarrow$ & SSIM $\uparrow$ & PSNR $\uparrow$
& nRMSE $\downarrow$ & SSIM $\uparrow$ & PSNR $\uparrow$
& nRMSE $\downarrow$ & SSIM $\uparrow$ & PSNR $\uparrow$ \\
\midrule

\rowcolor{datasetshade}
\multicolumn{13}{l}{\textit{Forward Prediction}} \\
DPOT-M
& \cellcolor{rankthird}0.0668 & \cellcolor{rankthird}0.9583 & \cellcolor{rankthird}42.6563
& 0.3114 & 0.9740 & 31.3651
& \cellcolor{ranksecond}0.1298 & \cellcolor{ranksecond}0.8754 & \cellcolor{ranksecond}24.2102
& \cellcolor{ranksecond}0.0457 & \cellcolor{ranksecond}0.9941 & \cellcolor{ranksecond}49.3510 \\

Poseidon-B
& 0.0723 & 0.9557 & 41.7776
& 0.3113 & 0.9678 & 31.5130
& 0.1814 & 0.7655 & 20.4111
& 0.0694 & 0.9821 & 44.9689 \\

PDE-Transformer-B
& 0.0762 & 0.9502 & 41.4816
& 0.2744 & 0.9776 & 34.8036
& 0.1376 & \cellcolor{rankthird}0.8616 & 23.7399
& \cellcolor{rankthird}0.0544 & 0.9881 & \cellcolor{rankthird}47.9531 \\

PDE-Transformer-FM
& 0.0873 & 0.9354 & 39.5308
& 0.2228 & 0.9814 & 38.8247
& \cellcolor{rankthird}0.1352 & 0.8103 & 22.5628
& 0.0920 & 0.9800 & 44.3802 \\

DYffusion
& 0.0792 & 0.9388 & 38.9746
& 0.2197 & 0.9796 & 37.1178
& 0.2176 & 0.5181 & 16.3631
& 0.2071 & 0.9287 & 34.9430 \\

PDE-Refiner
& 0.0917 & 0.9270 & 38.6464
& \cellcolor{rankthird}0.2037 & \cellcolor{rankthird}0.9850 & \cellcolor{rankthird}40.5664
& 0.1437 & 0.8333 & \cellcolor{rankthird}24.1400
& 0.0686 & \cellcolor{rankthird}0.9891 & 46.7395 \\

DiffUNet$^{2}$-DR (Ours)
& \cellcolor{ranksecond}0.0596 & \cellcolor{ranksecond}0.9671 & \cellcolor{ranksecond}43.4184
& \cellcolor{ranksecond}0.1700 & \cellcolor{ranksecond}0.9885 & \cellcolor{rankfirst}\textbf{43.1781}
& 0.1781 & 0.7672 & 22.2925
& 0.0910 & 0.9867 & 47.6509 \\

DiffUNet$^{2}$-AR (Ours)
& \cellcolor{rankfirst}\textbf{0.0572} & \cellcolor{rankfirst}\textbf{0.9681} & \cellcolor{rankfirst}\textbf{43.4572}
& \cellcolor{rankfirst}\textbf{0.1639} & \cellcolor{rankfirst}\textbf{0.9893} & \cellcolor{ranksecond}42.8285
& \cellcolor{rankfirst}\textbf{0.1287} & \cellcolor{rankfirst}\textbf{0.8774} & \cellcolor{rankfirst}\textbf{26.8785}
& \cellcolor{rankfirst}\textbf{0.0337} & \cellcolor{rankfirst}\textbf{0.9957} & \cellcolor{rankfirst}\textbf{53.3716} \\

\midrule
\rowcolor{datasetshade}
\multicolumn{13}{l}{\textit{Backward Prediction}} \\
DPOT-M
& \cellcolor{rankthird}0.0449 & \cellcolor{rankthird}0.9740 & \cellcolor{rankthird}44.3108
& 0.3122 & 0.9757 & 30.5255
& 0.2641 & 0.6404 & 15.3876
& 0.0996 & 0.9673 & 37.7370 \\

Poseidon-B
& 0.0475 & 0.9697 & 43.0327
& 0.3250 & 0.9709 & 30.1591
& 0.2842 & 0.5349 & 13.7300
& 0.0904 & 0.9738 & 39.9814 \\

PDE-Transformer-B
& 0.0490 & 0.9671 & 43.4721
& 0.2776 & 0.9706 & 31.5498
& 0.2953 & 0.5549 & 13.6719
& \cellcolor{ranksecond}0.0635 & 0.9754 & \cellcolor{rankthird}43.4501 \\

PDE-Transformer-FM
& 0.0455 & 0.9728 & 44.1173
& 0.2634 & \cellcolor{rankthird}0.9779 & 32.0348
& 0.2194 & 0.7035 & 15.8493
& 0.0760 & 0.9776 & 42.1634 \\

DYffusion
& 0.0807 & 0.9611 & 42.5716
& \cellcolor{rankthird}0.2568 & 0.9514 & \cellcolor{rankthird}32.2638
& \cellcolor{rankthird}0.2153 & 0.6025 & 15.7428
& 0.0924 & 0.9726 & 38.9524 \\

PDE-Refiner
& 0.0583 & 0.9652 & 42.1008
& 0.2875 & 0.9764 & 31.4011
& 0.2250 & \cellcolor{rankthird}0.7125 & \cellcolor{rankthird}16.4630
& 0.0686 & \cellcolor{rankthird}0.9905 & 43.2520 \\

DiffUNet$^{2}$-DR (Ours)
& \cellcolor{rankfirst}\textbf{0.0363} & \cellcolor{rankfirst}\textbf{0.9829} & \cellcolor{rankfirst}\textbf{46.1006}
& \cellcolor{rankfirst}\textbf{0.2215} & \cellcolor{rankfirst}\textbf{0.9847} & \cellcolor{rankfirst}\textbf{33.6656}
& \cellcolor{rankfirst}\textbf{0.1671} & \cellcolor{rankfirst}\textbf{0.8163} & \cellcolor{rankfirst}\textbf{18.3166}
& \cellcolor{rankthird}0.0637 & \cellcolor{ranksecond}0.9927 & \cellcolor{ranksecond}44.5622 \\

DiffUNet$^{2}$-AR (Ours)
& \cellcolor{ranksecond}0.0365 & \cellcolor{ranksecond}0.9827 & \cellcolor{ranksecond}45.6609
& \cellcolor{ranksecond}0.2371 & \cellcolor{ranksecond}0.9841 & \cellcolor{ranksecond}33.2329
& \cellcolor{ranksecond}0.2035 & \cellcolor{ranksecond}0.7304 & \cellcolor{ranksecond}17.8330
& \cellcolor{rankfirst}\textbf{0.0402} & \cellcolor{rankfirst}\textbf{0.9960} & \cellcolor{rankfirst}\textbf{48.7623} \\

\bottomrule
\end{tabular}
}
\end{table*}

%% file: table/table_modelcompare.tex
\begin{table}[t]
\centering
\caption{Evaluated temporal surrogate models and computational cost.
Type denotes deterministic/probabilistic prediction, Mode the temporal
prediction strategy, Format the supported data dimensionality, and Bidir.
whether forward and backward directions are trained separately or jointly.
Parameter counts are in millions.
Average training time is computed over the four 2D datasets; for separate
models, forward and backward training times are summed, while joint models
require one training run.
Average evaluation time is reported per predicted frame.
}
\label{tab:eval-baselines}
\fontsize{6.5}{7.5}\selectfont
\setlength{\tabcolsep}{1.2pt}
\renewcommand{\arraystretch}{0.90}
\setlength{\aboverulesep}{1pt}
\setlength{\belowrulesep}{1pt}
\begin{tabular*}{\columnwidth}{@{\extracolsep{\fill}}llllcrrr@{}}
\toprule
Method & Type & Mode & Format & Bidir. &
\shortstack{Params\\(M)} &
\shortstack{Avg. train\\(h)} &
\shortstack{Avg. eval.\\(s/frame)} \\
\midrule
DPOT-M~\cite{hao2024dpot}
& Det. & Rollout & 2D/3D & Separate
& 112.69 & 1.77 & 0.073 \\

Poseidon-B~\cite{herde2024poseidon}
& Det. & Rollout & 2D & Separate
& 157.73 & 2.96  & 0.133 \\

PDE-Transformer-B~\cite{holzschuh2025pdetransformer}
& Det. & Rollout & 2D & Separate
& 178.97 & 3.74 & 0.112 \\

PDE-Transformer-FM~\cite{holzschuh2025pdetransformer,lipman2023flowmatching}
& Prob. & Rollout & 2D & Separate
& 178.97 & 11.96  & 0.305 \\

PDE-Refiner~\cite{lippe2023pderefiner}
& Prob. & Rollout & 2D & Separate
& 148.69 & 2.72 & 0.102 \\

DYffusion~\cite{cachay2023dyffusion}
& Prob. & Direct & 2D & Separate
& 20.54 & 6.18 & 0.037 \\

DiffUNet$^2$-DR (Ours)
& Prob. & Direct & 2D/3D & Joint
& 57.78 & 4.85  & 0.193 \\

DiffUNet$^2$-AR (Ours)
& Prob. & Rollout & 2D/3D & Joint
& 56.73 & 4.03  & 0.197 \\
\bottomrule
\end{tabular*}
\end{table}

%% file: table/table2.tex
\definecolor{datasetshade}{RGB}{245,245,245}
\begin{table}[t]
\centering
\caption{
Distributional quality and spatial uncertainty--error correspondence
using 32 generated samples. Lower CRPS is better; ideal SSR is one.
$\rho$ is the Spearman correlation between ensemble standard deviation
and absolute ensemble-mean error, averaged within each trajectory
and then across trajectories. Higher $\rho$ is better.
Best results are bold.
}
\label{tab:distribution_quality}
\fontsize{4}{4}\selectfont
\setlength{\tabcolsep}{5.5pt}
\renewcommand{\arraystretch}{0.90}
\setlength{\heavyrulewidth}{0.1em}
\setlength{\lightrulewidth}{0.1em}
\setlength{\cmidrulewidth}{0.1em}
\setlength{\aboverulesep}{0.3pt}
\setlength{\belowrulesep}{1pt}
\resizebox{\columnwidth}{!}{
\begin{tabular}{lccc ccc}
\toprule
\textbf{Methods}
& \multicolumn{3}{c}{\textbf{Forward Prediction}}
& \multicolumn{3}{c}{\textbf{Backward Prediction}} \\
\cmidrule(lr){2-4}
\cmidrule(lr){5-7}
& CRPS $\downarrow$ & SSR $\rightarrow 1$ & $\rho$ $\uparrow$
& CRPS $\downarrow$ & SSR $\rightarrow 1$ & $\rho$ $\uparrow$ \\
\midrule
\rowcolor{datasetshade}
\multicolumn{7}{l}{\textit{CERM}} \\
PDE-Transformer-FM & 0.0284 & 0.7283 & 0.4489 & 0.0172 & \textbf{0.7737} & 0.4211 \\
DYffusion & 0.0308 & 0.6019 & 0.4158 & 0.0246 & 0.4614 & 0.3972 \\
PDE-Refiner & 0.0308 & 0.4806 & \textbf{0.5258} & 0.0194 & 0.4910 & \textbf{0.4916} \\
DiffUNet$^{2}$-DR (Ours) & 0.0195 & 0.5614 & 0.4562 & \textbf{0.0136} & 0.6111 & 0.4636 \\
DiffUNet$^{2}$-AR (Ours) & \textbf{0.0184} & \textbf{0.7807} & 0.4539 & 0.0141 & 0.5911 & 0.4323 \\
\addlinespace[1pt]
\rowcolor{datasetshade}
\multicolumn{7}{l}{\textit{Heat}} \\
PDE-Transformer-FM & 0.0475 & 0.6757 & 0.4293 & 0.0601 & 0.5916 & 0.6674 \\
DYffusion & 0.0634 & \textbf{0.8345} & \textbf{0.7306} & 0.0806 & \textbf{0.7660} & \textbf{0.8366} \\
PDE-Refiner & 0.0439 & 0.4428 & 0.2415 & 0.0732 & 0.3532 & 0.5472 \\
DiffUNet$^{2}$-DR (Ours) & 0.0312 & 0.5188 & 0.2896 & \textbf{0.0437} & 0.4260 & 0.2235 \\
DiffUNet$^{2}$-AR (Ours) & \textbf{0.0288} & 0.5315 & 0.1917 & 0.0466 & 0.3977 & 0.2479 \\
\addlinespace[1pt]
\rowcolor{datasetshade}
\multicolumn{7}{l}{\textit{Gray--Scott}} \\
PDE-Transformer-FM & 0.0282 & \textbf{1.0322} & \textbf{0.6936} & 0.0506 & 0.5321 & 0.6818 \\
DYffusion & 0.0540 & 0.8834 & 0.5146 & 0.0489 & \textbf{0.7818} & 0.6733 \\
PDE-Refiner & 0.0280 & 0.7001 & 0.6369 & 0.0532 & 0.3721 & 0.6987 \\
DiffUNet$^{2}$-DR (Ours) & 0.0415 & 0.3488 & 0.6089 & \textbf{0.0370} & 0.2556 & 0.6820 \\
DiffUNet$^{2}$-AR (Ours) & \textbf{0.0222} & 0.7222 & 0.6808 & 0.0455 & 0.4570 & \textbf{0.6991} \\
\addlinespace[1pt]
\rowcolor{datasetshade}
\multicolumn{7}{l}{\textit{NS-SVS}} \\
PDE-Transformer-FM & 0.0047 & 0.5802 & 0.6057 & 0.0045 & 0.4643 & 0.6056 \\
DYffusion & 0.0111 & \textbf{1.0140} & \textbf{0.6210} & 0.0059 & \textbf{0.8471} & 0.5461 \\
PDE-Refiner & 0.0034 & 0.5769 & 0.5624 & 0.0035 & 0.3234 & \textbf{0.6102} \\
DiffUNet$^{2}$-DR (Ours) & 0.0041 & 0.3144 & 0.5917 & 0.0029 & 0.3330 & 0.6054 \\
DiffUNet$^{2}$-AR (Ours) & \textbf{0.0017} & 0.6746 & 0.4686 & \textbf{0.0018} & 0.3807 & 0.5272 \\
\bottomrule
\end{tabular}
}
\end{table}

%% file: table/table3.tex
\definecolor{datasetshade}{RGB}{245,245,245}
\begin{table}[t]
\centering
\caption{
Ablation results. 
We compare separate training, joint bidirectional training, and joint training
with ambiguity-preserving cycle consistency for DiffUNet$^2$-DR and
DiffUNet$^2$-AR.
Fwd/Bwd denote forward/backward prediction.
Lower nRMSE and CRPS are better. Best results within each DR/AR group are bold.
}
\label{tab:ablation}

\fontsize{6}{7.2}\selectfont
\setlength{\tabcolsep}{1pt}
\renewcommand{\arraystretch}{0.95}
\setlength{\aboverulesep}{1pt}
\setlength{\belowrulesep}{1pt}

\begin{tabular*}{\columnwidth}
{@{\extracolsep{\fill}}l*{8}{c}@{}}
\toprule
& \multicolumn{2}{c}{CERM}
& \multicolumn{2}{c}{HEAT}
& \multicolumn{2}{c}{Gray--Scott}
& \multicolumn{2}{c}{NS-SVS} \\
\cmidrule(lr){2-3}\cmidrule(lr){4-5}
\cmidrule(lr){6-7}\cmidrule(lr){8-9}
Setting & Fwd & Bwd & Fwd & Bwd & Fwd & Bwd & Fwd & Bwd \\
\midrule
\rowcolor{datasetshade}
\multicolumn{9}{l}{\textit{nRMSE $\downarrow$}} \\
DR Separate
& 0.0592 & 0.0500 & \textbf{0.1580} & 0.2345
& 0.1862 & 0.1813 & \textbf{0.0807} & \textbf{0.0630} \\
DR Joint
& \textbf{0.0585} & 0.0389 & 0.1598 & 0.2350
& 0.1948 & 0.1791 & 0.0864 & 0.0647 \\
DR Joint + Cycle
& 0.0596 & \textbf{0.0363} & 0.1700 & \textbf{0.2215}
& \textbf{0.1781} & \textbf{0.1671} & 0.0910 & 0.0637 \\
\midrule[0.1pt]%\addlinespace[2pt]
AR Separate
& 0.0647 & 0.0427 & 0.1687 & 0.2358
& 0.1285 & \textbf{0.1927} & \textbf{0.0302} & 0.0454 \\
AR Joint
& 0.0575 & 0.0368 & 0.1655 & \textbf{0.2344}
& \textbf{0.1273} & 0.2028 & 0.0349 & 0.0413 \\
AR Joint + Cycle
& \textbf{0.0572} & \textbf{0.0365} & \textbf{0.1639} & 0.2371
& 0.1287 & 0.2035 & 0.0337 & \textbf{0.0402} \\
\midrule
\rowcolor{datasetshade}
\multicolumn{9}{l}{\textit{CRPS $\downarrow$}} \\
DR Separate
& 0.0209 & 0.0148 & \textbf{0.0289} & 0.0485
& 0.0446 & 0.0433 & \textbf{0.0037} & \textbf{0.0028} \\
DR Joint
& \textbf{0.0194} & 0.0140 & 0.0290 & 0.0482
& 0.0490 & 0.0416 & 0.0040 & 0.0030 \\
DR Joint + Cycle
& 0.0195 & \textbf{0.0136} & 0.0312 & \textbf{0.0437}
& \textbf{0.0415} & \textbf{0.0370} & 0.0041 & 0.0029 \\
\midrule[0.1pt]%\addlinespace[2pt]
AR Separate
& 0.0205 & 0.0155 & 0.0307 & 0.0472
& 0.0223 & \textbf{0.0422} & \textbf{0.0016} & 0.0020 \\
AR Joint
& 0.0185 & 0.0142 & 0.0294 & \textbf{0.0461}
& \textbf{0.0219} & 0.0451 & 0.0017 & \textbf{0.0018} \\
AR Joint + Cycle
& \textbf{0.0184} & \textbf{0.0141} & \textbf{0.0288} & 0.0466
& 0.0222 & 0.0455 & 0.0017 & 0.0018 \\
\bottomrule
\end{tabular*}
\end{table}

%% file: template.bib
@article{schneider2017climate,
    author  = {Schneider, Tapio and Teixeira, Jo{\~a}o and Bretherton, Christopher S. and others},
    title   = {Climate goals and computing the future of clouds},
    journal = {Nature Climate Change},
    volume  = {7},
    number  = {1},
    pages   = {3--5},
    year    = {2017},
    doi     = {10.1038/nclimate3190},
    url     = {https://doi.org/10.1038/nclimate3190},
    month   = jan,
    publisher = {Nature Publishing Group}
}

@article{Anirudh2020,
    author = {Rushil Anirudh  and Jayaraman J. Thiagarajan  and Peer-Timo Bremer  and Brian K. Spears },
    title = {Improved surrogates in inertial confinement fusion with manifold and cycle consistencies},
    journal = {Proceedings of the National Academy of Sciences},
    volume = {117},
    number = {18},
    pages = {9741-9746},
    year = {2020},
    doi = {10.1073/pnas.1916634117},
    URL = {https://www.pnas.org/doi/abs/10.1073/pnas.1916634117},
    eprint = {https://www.pnas.org/doi/pdf/10.1073/pnas.1916634117}
}

@inproceedings{ho2020ddpm,
    author = {Ho, Jonathan and Jain, Ajay and Abbeel, Pieter},
    title = {Denoising diffusion probabilistic models},
    year = {2020},
    isbn = {9781713829546},
    booktitle = {Proceedings of the 34th International Conference on Neural Information Processing Systems},
    series = {NIPS '20},
    doi={https://doi.org/10.48550/arXiv.2006.11239}
}

@misc{nichol2021iddpm,
    title={Improved Denoising Diffusion Probabilistic Models}, 
    author={Alex Nichol and Prafulla Dhariwal},
    year={2021},
    eprint={2102.09672},
    archivePrefix={arXiv},
    primaryClass={cs.LG},
    url={https://arxiv.org/abs/2102.09672}, 
    doi={https://doi.org/10.48550/arXiv.2102.09672}
}

@misc{song2022ddim,
    title={Denoising Diffusion Implicit Models}, 
    author={Jiaming Song and Chenlin Meng and Stefano Ermon},
    year={2022},
    eprint={2010.02502},
    archivePrefix={arXiv},
    primaryClass={cs.LG},
    url={https://arxiv.org/abs/2010.02502},
    doi={https://doi.org/10.48550/arXiv.2010.02502}
}

@misc{li2021fourier,
      title={Fourier Neural Operator for Parametric Partial Differential Equations}, 
      author={Zongyi Li and Nikola Kovachki and Kamyar Azizzadenesheli and Burigede Liu and Kaushik Bhattacharya and Andrew Stuart and Anima Anandkumar},
      year={2021},
      eprint={2010.08895},
      archivePrefix={arXiv},
      primaryClass={cs.LG},
      url={https://arxiv.org/abs/2010.08895}, 
      doi={https://doi.org/10.48550/arXiv.2010.08895}
}

@inproceedings{takamoto2022pdebench,
    author = {Takamoto, Makoto and Praditia, Timothy and Leiteritz, Raphael and MacKinlay, Dan and Alesiani, Francesco and Pfl\"{u}ger, Dirk and Niepert, Mathias},
    title = {PDEBENCH: an extensive benchmark for scientific machine learning},
    year = {2022},
    isbn = {9781713871088},
    publisher = {Curran Associates Inc.},
    address = {Red Hook, NY, USA},
    booktitle = {Proceedings of the 36th International Conference on Neural Information Processing Systems},
    articleno = {117},
    numpages = {16},
    location = {New Orleans, LA, USA},
    series = {NIPS '22},
    doi={https://doi.org/10.48550/arXiv.2210.07182}
}

@article{lam2023learning,
    title={Learning skillful medium-range global weather forecasting},
    author={Lam, Remi and Sanchez-Gonzalez, Alvaro and Willson, Matthew and Wirnsberger, Peter and Fortunato, Meire and Alet, Ferran and Ravuri, Suman and Ewalds, Timo and Eaton-Rosen, Zach and Hu, Weihua and others},
    journal={Science},
    volume={382},
    number={6677},
    pages={1416--1421},
    year={2023},
    publisher={American Association for the Advancement of Science},
    doi={https://doi.org/10.1126/science.adi2336}
}

@article{liu2019solar,
    url = {https://doi.org/10.3847/1538-4357/ab1b3c},
    year = {2019},
    month = {jun},
    publisher = {The American Astronomical Society},
    volume = {877},
    number = {2},
    pages = {121},
    author = {Liu, Hao and Liu, Chang and Wang, Jason T. L. and Wang, Haimin},
    title = {Predicting Solar Flares Using a Long Short-term Memory Network},
    journal = {The Astrophysical Journal},
    doi = {10.3847/1538-4357/ab1b3c}
}

@article{li2024SEEDS,
    title	= {Generative emulation of weather forecast ensembles with diffusion models},
    author={Li, Lizao and Carver, Robert and Lopez-Gomez, Ignacio and Sha, Fei and Anderson, John},
    year	= {2024},
    URL	= {https://www.science.org/doi/10.1126/sciadv.adk4489},
    journal	= {Science Advances},
    pages	= {eadk4489},
    volume	= {10},
    doi={https://doi.org/10.1126/sciadv.adk4489}
}

@inproceedings{andrae2025continuous,
    title={Continuous Ensemble Weather Forecasting with Diffusion models},
    author={Martin Andrae and Tomas Landelius and Joel Oskarsson and Fredrik Lindsten},
    booktitle={The Thirteenth International Conference on Learning Representations},
    year={2025},
    url={https://openreview.net/forum?id=ePEZvQNFDW},
    doi={https://doi.org/10.48550/arXiv.2410.05431}
}

@misc{shu2024zeroshot,
    title={Zero-Shot Uncertainty Quantification using Diffusion Probabilistic Models}, 
    author={Dule Shu and Amir Barati Farimani},
    year={2024},
    eprint={2408.04718},
    archivePrefix={arXiv},
    primaryClass={cs.LG},
    url={https://arxiv.org/abs/2408.04718}, 
    doi={https://doi.org/10.48550/arXiv.2408.04718}
}

@article{kingma2019VAE,
    author = {Kingma, Diederik P. and Welling, Max},
    title = {An Introduction to Variational Autoencoders},
    journal = {Foundations and Trends in Machine Learning},
    volume = {12},
    number = {4},
    pages = {307-392},
    year = {2019},
    month = {11},
    issn = {1935-8237},
    doi = {10.1561/2200000056},
    url = {https://doi.org/10.1561/2200000056},
    eprint = {https://www.emerald.com/ftmal/article-pdf/12/4/307/11160827/2200000056en.pdf},
}

@inproceedings{Goodfellow2014GAN,
    author = {Goodfellow, Ian J. and Pouget-Abadie, Jean and Mirza, Mehdi and Xu, Bing and Warde-Farley, David and Ozair, Sherjil and Courville, Aaron and Bengio, Yoshua},
    booktitle = {Advances in Neural Information Processing Systems},
    editor = {Z. Ghahramani and M. Welling and C. Cortes and N. Lawrence and K.Q. Weinberger},
    publisher = {Curran Associates, Inc.},
    title={Generative Adversarial Networks},
    url = {https://proceedings.neurips.cc/paper_files/paper/2014/file/f033ed80deb0234979a61f95710dbe25-Paper.pdf},
    volume = {27},
    year = {2014},
    doi={https://doi.org/10.48550/arXiv.1406.2661}
}

@inproceedings{cachay2023dyffusion,
    author = {R\"{u}hling Cachay, Salva and Zhao, Bo and Joren, Hailey and Yu, Rose},
    booktitle = {Advances in Neural Information Processing Systems},
    editor = {A. Oh and T. Naumann and A. Globerson and K. Saenko and M. Hardt and S. Levine},
    pages = {45259--45287},
    publisher = {Curran Associates, Inc.},
    title = {DYffusion: A Dynamics-informed Diffusion Model for Spatiotemporal Forecasting},
    url = {https://proceedings.neurips.cc/paper_files/paper/2023/file/8df90a1440ce782d1f5607b7a38f2531-Paper-Conference.pdf},
    volume = {36},
    year = {2023},
    doi = {https://doi.org/10.48550/arXiv.2306.01984}
}

@article{yu2026probabilistic,
    AUTHOR = {Yu, W. and Ghosh, A. and Finn, T. S. and Arcucci, R. and Bocquet, M. and Cheng, S.},
    TITLE = {A probabilistic approach to wildfire spread prediction using a denoising diffusion surrogate model},
    JOURNAL = {Geoscientific Model Development},
    VOLUME = {19},
    YEAR = {2026},
    NUMBER = {2},
    PAGES = {1027--1054},
    URL = {https://gmd.copernicus.org/articles/19/1027/2026/},
    DOI = {10.5194/gmd-19-1027-2026}
}

@article{Banesh2025HEATPLI,
    title = {The High Explosives \& Affected Targets (HEAT) Dataset},
    author = {Banesh, Divya and Chakrabarti, Sharmistha and De, Soumi and Egozi, Gal and Hickmann, Kyle and Hlavacek, William S. and Kaiser, Bryan and Pandit, Sourabh and Pulido, Jesus and Schodt, David and Sweeney, Christine},
    abstractNote = {Physics-rich dataset of cylindrically symmetric shock-propagation simulations (CYL and PLI).},
    doi = {10.25583/2571471},
    place = {United States},
    year = {2025},
    month = {7},
    journal = {}
}

@article{raissi2019physics,
    title = {Physics-informed neural networks: A deep learning framework for solving forward and inverse problems involving nonlinear partial differential equations},
    journal = {Journal of Computational Physics},
    volume = {378},
    pages = {686-707},
    year = {2019},
    issn = {0021-9991},
    doi = {https://doi.org/10.1016/j.jcp.2018.10.045},
    url = {https://www.sciencedirect.com/science/article/pii/S0021999118307125},
    author = {M. Raissi and P. Perdikaris and G.E. Karniadakis}
}

@article{moreland2013pipe,
    author={Moreland, Kenneth},
    journal={IEEE Transactions on Visualization and Computer Graphics}, 
    title={A Survey of Visualization Pipelines}, 
    year={2013},
    volume={19},
    number={3},
    pages={367-378},
    doi={10.1109/TVCG.2012.133}
}

@misc{casiraghi2024complexsystem,
    title={Disentangling the Timescales of a Complex System: A Bayesian Approach to Temporal Network Analysis}, 
    author={Giona Casiraghi and Georges Andres},
    year={2024},
    eprint={2403.05343},
    archivePrefix={arXiv},
    primaryClass={stat.ME},
    url={https://arxiv.org/abs/2403.05343}, 
    doi={https://doi.org/10.48550/arXiv.2403.05343}
}

@article{yitang2025INR,
    author={Chen, Yi-Tang and Li, Haoyu and Shi, Neng and Luo, Xihaier and Xu, Wei and Shen, Han-Wei},
    journal={ IEEE Transactions on Visualization \& Computer Graphics },
    title={{ Explorable INR: An Implicit Neural Representation for Ensemble Simulation Enabling Efficient Spatial and Parameter Exploration }},
    year={2025},
    volume={31},
    number={06},
    ISSN={1941-0506},
    pages={3758-3770},
    doi={10.1109/TVCG.2025.3567052},
    url = {https://doi.ieeecomputersociety.org/10.1109/TVCG.2025.3567052},
    publisher={IEEE Computer Society},
    address={Los Alamitos, CA, USA},
    month=jun
}

@book{strogatz2018nonlinear,
    author = {Strogatz, Steven H},
    title = {Nonlinear Dynamics and Chaos: With Applications to Physics, Biology, Chemistry, and Engineering (3rd ed.)},
    publisher = {Chapman and Hall/CRC.},
    year = {2024},
    doi={https://doi.org/10.1201/9780429398490}
}

@inproceedings{song2021score,
    author = {Yang Song and Jascha Sohl{-}Dickstein and Diederik P. Kingma and Abhishek Kumar and Stefano Ermon and Ben Poole},
    title = {Score-Based Generative Modeling through Stochastic Differential Equations},
    booktitle = {9th International Conference on Learning Representations, {ICLR} 2021, Virtual Event, Austria, May 3-7, 2021},
    publisher    = {OpenReview.net},
    year         = {2021},
    url          = {https://openreview.net/forum?id=PxTIG12RRHS},
    bibsource    = {dblp computer science bibliography, https://dblp.org},
    doi={https://doi.org/10.48550/arXiv.2011.13456}
}

@inproceedings{hao2024dpot,
  title = {{DPOT}: Auto-Regressive Denoising Operator Transformer for Large-Scale {PDE} Pre-Training},
  author = {Hao, Zhongkai and Su, Chang and Liu, Songming and Berner, Julius and Ying, Chengyang and Su, Hang and Anandkumar, Anima and Song, Jian and Zhu, Jun},
  booktitle = {Proceedings of the 41st International Conference on Machine Learning},
  volume = {235}, pages = {17616--17635}, year = {2024},
  editor = 	 {Salakhutdinov, Ruslan and Kolter, Zico and Heller, Katherine and Weller, Adrian and Oliver, Nuria and Scarlett, Jonathan and Berkenkamp, Felix},
  series = 	 {Proceedings of Machine Learning Research},
  month = 	 {21--27 Jul},
  publisher =    {PMLR},
  url = 	 {https://proceedings.mlr.press/v235/hao24d.html},
  doi = {https://doi.org/10.48550/arXiv.2403.03542}
}

@inproceedings{herde2024poseidon,
  title = {{Poseidon}: Efficient Foundation Models for {PDEs}},
  author = {Herde, Maximilian and Raoni\'{c}, Bogdan and Rohner, Tobias and K\"{a}ppeli, Roger and Molinaro, Roberto and de B\'{e}zenac, Emmanuel and Mishra, Siddhartha},
  booktitle = {Advances in Neural Information Processing Systems},
  doi = {10.52202/079017-2311},
  editor = {A. Globerson and L. Mackey and D. Belgrave and A. Fan and U. Paquet and J. Tomczak and C. Zhang},
  pages = {72525--72624},
  publisher = {Curran Associates, Inc.},
  url = {https://proceedings.neurips.cc/paper_files/paper/2024/file/84e1b1ec17bb11c57234e96433022a9a-Paper-Conference.pdf},
  volume = {37},
  year = {2024}
}

@inproceedings{holzschuh2025pdetransformer,
    title={{PDE}-Transformer: Efficient and Versatile Transformers for Physics Simulations},
    author={Benjamin Holzschuh and Qiang Liu and Georg Kohl and Nils Thuerey},
    booktitle={Forty-second International Conference on Machine Learning, {ICML} 2025, Vancouver, Canada},
    year={2025},
    url={https://openreview.net/forum?id=3BaJMRaPSx},
    doi={https://doi.org/10.48550/arXiv.2505.24717}
}

@inproceedings{lipman2023flowmatching,
  title = {Flow Matching for Generative Modeling},
  author = {Lipman, Yaron and Chen, Ricky T. Q. and Ben-Hamu, Heli and Nickel, Maximilian and Le, Matt},
  booktitle = {International Conference on Learning Representations}, year = {2023},
  url = {https://arxiv.org/abs/2210.02747},
  doi = {https://doi.org/10.48550/arXiv.2210.02747}
}

@inproceedings{lippe2023pderefiner,
    author = {Lippe, Phillip and Veeling, Bastiaan S. and Perdikaris, Paris and Turner, Richard E. and Brandstetter, Johannes},
    title = {PDE-refiner: achieving accurate long rollouts with neural PDE solvers},
    year = {2023},
    booktitle = {Proceedings of the 37th International Conference on Neural Information Processing Systems},
    articleno = {2946},
    numpages = {36},
    location = {New Orleans, LA, USA},
    series = {NIPS '23},
    doi = {https://doi.org/10.48550/arXiv.2308.05732}
}

@article{pearson1993patterns,
  title = {Complex Patterns in a Simple System},
  author = {Pearson, John E.},
  journal = {Science}, volume = {261}, number = {5118}, pages = {189--192},
  year = {1993}, doi = {10.1126/science.261.5118.189}
}

@inproceedings{koehler2024apebench,
  title = {{APEBench}: A Benchmark for Autoregressive Neural Emulators of {PDEs}},
  author = {Koehler, Felix and Niedermayr, Simon and Westermann, R{\"u}diger and Thuerey, Nils},
  booktitle = {Advances in Neural Information Processing Systems},
  publisher = {Curran Associates, Inc.},
  doi = {10.52202/079017-3822},
  year = {2024},
  url ={https://proceedings.neurips.cc/paper_files/paper/2024/file/d9875ebcf74bccdc5076acab0dbee62c-Paper-Datasets_and_Benchmarks_Track.pdf},
  volume = {37}
}

@article{stuart2010inverse,
  author  = {Stuart, A. M.},
  title   = {Inverse Problems: A Bayesian Perspective},
  journal = {Acta Numerica},
  volume  = {19},
  pages   = {451--559},
  year    = {2010},
  doi     = {10.1017/S0962492910000061}
}

@article{min2014backwardheat,
  author  = {Min, Tao and Fu, Weimin and Huang, Qiang},
  title   = {Inverse Estimates for Nonhomogeneous Backward Heat Problems},
  journal = {Journal of Applied Mathematics},
  volume  = {2014},
  pages   = {529618},
  year    = {2014},
  doi     = {10.1155/2014/529618}
}

@article{luo2023cfdbench,
  title   = {{CFDBench}: A Large-Scale Benchmark for Machine Learning Methods in Fluid Dynamics},
  author  = {Luo, Yining and Chen, Yingfa and Zhang, Zhen},
  journal = {arXiv preprint arXiv:2310.05963},
  year    = {2023},
  doi = {
https://doi.org/10.48550/arXiv.2310.05963}
}

@article{sun2025prosepde,
  title = {Towards a foundation model for partial differential equations: Multioperator learning and extrapolation},
  author = {Sun, Jingmin and Liu, Yuxuan and Zhang, Zecheng and Schaeffer, Hayden},
  journal = {Phys. Rev. E},
  volume = {111},
  issue = {3},
  pages = {035304},
  numpages = {18},
  year = {2025},
  month = {Mar},
  publisher = {American Physical Society},
  doi = {10.1103/PhysRevE.111.035304},
  url = {https://link.aps.org/doi/10.1103/PhysRevE.111.035304}
}

@inproceedings{molinaro2023neuralinverse,
    author = {Molinaro, Roberto and Yang, Yunan and Engquist, Bj{\"o}rn and Mishra, Siddhartha},
    title = {Neural inverse operators for solving PDE inverse problems},
    year = {2023},
    publisher = {JMLR.org},
    booktitle = {Proceedings of the 40th International Conference on Machine Learning},
    articleno = {1046},
    numpages = {35},
    location = {Honolulu, Hawaii, USA},
    series = {ICML'23},
    doi = {https://dl.acm.org/doi/10.5555/3618408.3619454}
}

@inproceedings{long2025invertible,
  title     = {Invertible Fourier Neural Operators for Tackling Both Forward and Inverse Problems},
  author    = {Long, Da and Xu, Zhitong and Yuan, Qiwei and Yang, Yin and Zhe, Shandian},
  booktitle = {Proceedings of the 28th International Conference on Artificial Intelligence and Statistics},
  volume    = {258},
  pages     = {3043--3051},
  year      = {2025},
  publisher = {PMLR},
  doi = {https://doi.org/10.48550/arXiv.2402.11722}
}

@inproceedings{haitsiukevich2024diffusion,
  author={Haitsiukevich, Katsiaryna and Poyraz, Onur and Marttinen, Pekka and Ilin, Alexander},
  booktitle={2024 IEEE 34th International Workshop on Machine Learning for Signal Processing (MLSP)}, 
  title={Diffusion Models as Probabilistic Neural Operators for Recovering Unobserved States of Dynamical Systems}, 
  year={2024},
  pages={1-6},
  doi={10.1109/MLSP58920.2024.10734762}
}
